\documentclass[12pt]{article}

\ifx\pdfoutput\undefined
\usepackage[dvips,bookmarks]{hyperref}
\else
\usepackage{hyperref}
\fi
\hypersetup{colorlinks=false,bookmarksopen,bookmarksnumbered,citecolor=blue,
   pdfstartview=FitH}

\usepackage{latexsym}
\usepackage{amssymb,amsfonts,amsmath}
\usepackage{graphicx} 
\usepackage{indentfirst}
\usepackage{bbm}
\parskip=\medskipamount

\newcommand {\cD}{{\cal D}}
\newcommand {\cE}{{\cal E}}

\newcommand {\cG}{{\cal G}}
\newcommand {\cH}{{\cal H}}

\newcommand {\cJ}{{\cal J}}

\newcommand {\cL}{{\cal L}}
\newcommand {\cM}{{\cal M}}
\newcommand {\cN}{{\cal N}}

\newcommand {\cT}{{\cal T}}
\newcommand {\cU}{{\cal U}}
\newcommand {\cV}{{\cal V}}
\newcommand {\cW}{{\cal W}}

\newcommand {\cY}{{\cal Y}}

\def\a{\alpha}
\def\b{\beta}
\def\c{\chi}
\def\d{\delta}

\def\f{\phi}
\def\g{\gamma}
\def\G{\Gamma}

\def\k{\kappa}
\def\l{\lambda}

\def\q{\theta}
\def\r{\rho}
\def\s{\sigma}

\def\x{\xi}
\def\z{\zeta}
\def\D{\Delta}
\def\F{\Phi}
\def\J{\Psi}
\def\L{\Lambda}
\def\O{\Omega}

\def\S{\Sigma}
\def\U{\Upsilon}

\def\ri{{\rm i}}

\newcommand{\ad}{{\dot{\alpha}}}                           %new
\newcommand{\bd}{{\dot{\beta}}}                            %new
\newcommand{\ve}{\varepsilon}                            %new
\newcommand{\cDB}{{\bar\cD}}                            %new

\newcommand{\pa}{\partial}                           %new
\newcommand{\hf}{\frac12}
\newcommand{\vf}{\varphi}
\newcommand{\be}{\begin{equation}}
\newcommand{\ee}{\end{equation}}
\newcommand{\bea}{\begin{eqnarray}}
\newcommand{\eea}{\end{eqnarray}}
\newcommand{\non}{\nonumber}
\newcommand{\ba}{\begin{array}}
\newcommand{\ea}{\end{array}}

\newcommand{\1}{{\underline{1}}}
\newcommand{\2}{{\underline{2}}}

\newcommand{\bm}[1]{\mbox{\boldmath$#1$}}

\def\double #1{#1{\hbox{\kern-2pt $#1$}}}

\newcommand{\gd}{{\dot\g}}
\newcommand{\dd}{{\dot\d}}

\newcommand{\CD}{{\nabla}}

\newcommand{\teb}{{\bar{\theta}}}

\newcommand{\bsubeq}{\begin{subequations}}
\newcommand{\esubeq}{\end{subequations}}

\newcommand{\ul}{\underline}
\newcommand{\eps}{{\epsilon}}

\newcommand{\dalpha}{{\dot{\alpha}}}
\newcommand{\dbeta}{{\dot{\beta}}}
\newcommand{\dgamma}{{\dot{\gamma}}}

\newcommand{\btheta}{{\bar\theta}}
\newcommand{\N}{{\mathcal N}}
\newcommand{\eol}{\notag \\}
\newcommand{\rd}{\mathrm d}
\newcommand{\qW}{{\mathbf W}}
\newcommand{\qG}{{\mathbf G}}
\newcommand{\qH}{{\mathbf H}}
\newcommand{\qPsi}{{\mathbf \Psi}}
\newcommand{\qV}{{\mathbf V}}
\newcommand{\qOmega}{{\mathbf \Omega}}

\newcommand{\BCD}{{\bar\cD}}

\newcommand{\gD}{{\mathbb D}}
\newcommand{\Unew}{{U}}
\newcommand{\cUnew}{\cU}
\begin{document}

\begin{titlepage}
\begin{flushright}
April 2011\\
\end{flushright}
\vspace{5mm}

\begin{center}
{\Large \bf $\bm{\cN = 2}$  AdS supergravity and supercurrents }
\\ 
\end{center}

\begin{center}

{\bf Daniel Butter and  Sergei M. Kuzenko }

\footnotesize{
{\it School of Physics M013, The University of Western Australia\\
35 Stirling Highway, Crawley W.A. 6009, Australia}}  ~\\
\texttt{dbutter,\,kuzenko@cyllene.uwa.edu.au}\\
\vspace{2mm}

\end{center}
\vspace{5mm}

\begin{abstract}
\baselineskip=14pt
We consider the minimal off-shell formulation 
for four-dimensional $\cN=2$ supergravity with a cosmological term, 
in which the second compensator is an improved tensor multiplet.  We use it to derive 
a linearized supergravity action (and its dual versions) around the anti-de Sitter (AdS) background 
in terms of three $\cN=2$  off-shell multiplets: 
an unconstrained  scalar superfield,  vector and tensor multiplets. 
This allows us to deduce the structure of the supercurrent multiplet associated with  
those supersymmetric theories which naturally couple to the supergravity formulation chosen, 
with or without a cosmological term. Finally, our linearized $\cN=2$ AdS supergravity 
action is reduced to $\cN=1$ superspace. The result is a sum of two $\cN=1$ linearized actions 
describing (i) old minimal supergravity; and (ii) an off-shell  massless gravitino multiplet.
We also derive dual formulations for the massless $\cN=1$ gravitino multiplet in AdS.
As a by-product of our consideration, we derive the consistent supergravity extension of 
the $\cN=1$ supercurrent multiplet advocated recently by Komargodski and Seiberg.
\end{abstract}
\vspace{0.5cm}
\begin{flushright}
{\it Dedicated To The 50th Anniversary \\
Of The First Man In Space}
\end{flushright}

\vfill
\end{titlepage}

\newpage
\renewcommand{\thefootnote}{\arabic{footnote}}
\setcounter{footnote}{0}

\tableofcontents

%\pagestyle{plain}
%\pagenumbering{arabic}

% Equation numbers
\numberwithin{equation}{section}

%\allowdisplaybreaks

%===BEGIN DOCUMENT=============================================================
%\newpage
\section{Introduction}
One of the reasons to study linearized off-shell supergravity actions around exact
supergravity backgrounds is the possibility to generate consistent supercurrent
multiplets, that is supermultiplets containing the 
energy-momentum tensor and the supersymmetry current(s) \cite{FZ}.
In a recent paper \cite{n2_sugra} we have found the linearized superfield action 
of the minimal 4D $\cN=2$ supergravity with a tensor compensator  \cite{deWPV} 
around a Minkowski superspace background.
This has allowed us to construct a new $\cN=2$ supercurrent multiplet, in addition to those
proposed in the past  \cite{Sohnius,Stelle,KT}.
In the present paper, we will extend the main constructions of \cite{n2_sugra} 
to the case of $\cN=2$ supergravity with a cosmological term. 
We will heavily use some of the results of our work \cite{n2_sugra_tensor} which in turn built 
on the series of papers \cite{KLRT-M1,KLRT-M2,K-dual08,KT-M}  concerning the projective-superspace
formulation for general $\cN=2$ supergravity-matter couplings. 

A natural question to ask is the following: Is there anything interesting to be learnt from an extension  
of the results in  \cite{n2_sugra}  to the anti-de Sitter (AdS) case?  The answer is certainly  `Yes'
in the sense that AdS supercurrent multiplets are usually more restrictive than those corresponding to the 
Poincar\'e supersymmetry.  To clarify this point, we would like to discuss, in some detail, 
the situation in $\cN=1$ supersymmetry.

In the case of $\cN=1$ Poincar\'e supersymmetry, 
the most general form (see, e.g.,  \cite{MSW,K-var})  of a supercurrent multiplet is as follows:
\bea
&{\bar D}^{\ad}{J} _{\a \ad} = { \c}_\a  +{\rm i}\,\eta_\a +D_\a X~, 
\label{Poincare_supercurrent} \\
&{\bar D}_\ad {\c}_\a  = {\bar D}_\ad \eta_\a= {\bar D}_\ad {X}=0~, 
\qquad D^\a {\c}_\a - {\bar D}_\ad {\bar {\c}}^\ad
=D^\a {\eta}_\a - {\bar D}_\ad {\bar {\eta}}^\ad = 0~.
\non
\eea
Here $J_{\a \ad} = {\bar J}_{\a \ad} $ denotes the supercurrent, while the chiral superfields 
$\c_\a$, $\eta_\a$ and $X$ constitute the so-called multiplet of anomalies. 
Some of the superfields $\c_\a$, $\eta_\a$ and $X$
are actually absent for concrete models, and all of them vanish in the case of superconformal theories.
The three terms on the right of (\ref{Poincare_supercurrent}) emphasize the fact that there  
exist exactly three different linearized actions for minimal ($12+12$) supergravity, according to the 
classification given in \cite{GKP}, which are related by superfield duality transformations. 
The case  $\c_\a = \eta_\a =0$ 
describes the  Ferrara-Zumino multiplet \cite{FZ} which 
 corresponds to the old minimal formulation 
for $\cN=1$ supergravity \cite{WZ,old}. The choice  $X= \eta_\a =0$
corresponds to the new minimal supergravity \cite{new} (this supercurrent was 
studied in \cite{tensor}). Finally, the third choice $X=\c_\a  =0$
corresponds to the minimal  supergravity formulation
proposed in \cite{BGLP}.\footnote{Unlike the old minimal and the new minimal 
theories, this formulation is known at the linearized level only.}

If only one of the superfields $\c_\a$, $\eta_\a$ and $X$  in  (\ref{Poincare_supercurrent})
is zero, the supercurrent multiplet describes $16+16$ components. Of the three such supercurrents 
studied in \cite{K-var},  the most  interesting is the one advocated by Komargodski and Seiberg \cite{KS2}.
Their conservation law is 
\bea
 {\bar D}^\ad J_{\a \ad} =  \c_\a +D_\a X~,
 \qquad {\bar D}_\ad {\c}_\a  = {\bar D}_\ad {X}=0~,  
\qquad D^\a {\c}_\a - {\bar D}_\ad {\bar {\c}}^\ad =0~.
\label{S-multiplet}
\eea
Finally, the most general supercurrent multiplet with $20+20$ components, 
for which all  the superfields $\c_\a$, $\eta_\a$ and $X$  in  (\ref{Poincare_supercurrent})
are non-zero, is related to a linearized version of the non-minimal formulation for $\cN=1$ supergravity
\cite{non-min,SG}.
The non-minimal supercurrent  can be written in the form \cite{GGRS}   
\bea
{\bar D}^{\ad}{J}_{\a \ad} = -\frac{1}{4} {\bar D}^2 
\z_\a
-\frac{1}{4} \frac{n+1}{3n+1} D_\a 
{\bar D}_\bd {\bar \z}^\bd 
~, \qquad D_{(\a}\z_{\b )} =0~,
\label{conservation-non}
\eea
where $n$ is a real constant, $n\neq-1/3, 0$, pararmetrizing the different versions of 
non-minimal supergravity \cite{SG}. The constraint on $\z_\a$ in 
(\ref{conservation-non}) is solved by $\z_\a = D_\a Z$, for some complex superfield $Z$, 
and then eq. (\ref{conservation-non}) takes the form  (\ref{Poincare_supercurrent}).

Let us now turn to the case of $\cN=1$ AdS supersymmetry. To start with, we could try to generalize 
the conservation equation (\ref{Poincare_supercurrent})  by replacing the flat covariant derivatives
$D_A= (\pa_a, D_\a, {\bar D}^\ad)$ in (\ref{Poincare_supercurrent})
with those corresponding to the AdS superspace, 
$D_A \to \nabla_A =(\nabla_a, \nabla_\a, {\bar \nabla}^\ad)$.
However, a simple analysis shows that the only consistent generalization 
obtained in this way is 
\bea
{\bar \nabla}^{\ad}{J} _{\a \ad} = \nabla_\a X~, \qquad {\bar \nabla}_\ad X =0~.
\label{AdS-mcl}
\eea
It corresponds to the old minimal supergravity with a cosmological term, 
for which the linearized action around the AdS background  \cite{KS} is
\bea
S_{\rm old} &=& - \int \rd^4x\, \rd^2\theta\, \rd^2\btheta\, E\, \Big\{
      \frac{1}{16} H^{\dalpha \alpha} \nabla^\beta (\bar\nabla^2 - 4R) \nabla_\beta H_{\alpha \dalpha}
- \frac{1}{48} ([\nabla_\alpha, \bar\nabla_\dalpha] H^{\dalpha \alpha})^2 
     \non \\
     & +& 
     \frac{1}{4} (\nabla_{\alpha \dalpha} H^{\dalpha \alpha})^2
     + \frac{R \bar R}{4} H^{\dalpha \alpha} H_{\alpha \dalpha}
     + \ri H^{\dalpha \alpha} \nabla_{\alpha \dalpha} (\f - \bar \f)
     + 3 ( \f \bar\f  -  \f^2 -  \bar\f^2) \Big\}~,~~~~
\label{eq_N1OldMin}
\eea
with $H_{\a \ad}$ the gravitational superfield, 
$\f$ the chiral compensator, ${\bar \nabla}_\ad \f =0$, and
$R$ the constant torsion of the AdS superspace.
This action is invariant under the linearized supergravity gauge  transformations
\begin{align}\label{eq_N1gaugeOldMin}
\delta H_{\alpha \dalpha} &= \nabla_\alpha {\bar L}_\dalpha - \bar \nabla_\dalpha L_\alpha ~, \qquad
\delta \f = - \frac{1}{12} (\bar \nabla^2 - 4 R) \nabla^\alpha L_{\alpha}~,
\end{align}
with $L_\a$ an unconstrained superfield parameter.  Looking at the explicit structure
of the action (\ref{eq_N1OldMin}), it is easy to understand why the AdS supersymmetry allows only for 
one minimal supercurrent multiplet, which is given by eq. (\ref{AdS-mcl}).
In particular,
the Komargodski-Seiberg supercurrent  (\ref{S-multiplet}) does not admit a minimal AdS extension.
The point is that  the theory (\ref{eq_N1OldMin}) does not possess, for $R \neq 0$,  a dual formulation in which
the chiral compensator $\f$ and its conjugate $\bar \f$ get replaced by a real linear superfield. 

It is instructive to see how the consistency issue mentioned arises in terms of the non-minimal supercurrent 
(\ref{conservation-non}).  Starting from (\ref{conservation-non}), 
let us again replace the flat covariant derivatives
$D_A= (\pa_a, D_\a, {\bar D}^\ad)$
with the AdS ones, 
$D_A \to \nabla_A =(\nabla_a, \nabla_\a, {\bar \nabla}^\ad)$.
It turns out that such a generalization is consistent only in the case $n=-1$,
\bea
{\bar \nabla}^{\ad}{J}_{\a \ad} = -\frac{1}{4} {\bar \nabla}^2 
\z_\a ~, \qquad \nabla_{(\a}\z_{\b )} =0~.
\eea
This supercurrent multiplet is associated with the linearized supergravity action \cite{KS}
\bea
S_{n=-1} &=&- \int \rd^4x\, \rd^2\theta\, \rd^2\btheta\, E\, \Big\{
       \frac{1}{16} H^{\dalpha \alpha} \nabla^\beta (\bar\nabla^2 - 4R) \nabla_\beta H_{\alpha \dalpha} 
      + \frac{1}{4} R \bar R H^{\dalpha \alpha} H_{\alpha \dalpha}
      \non \\
&& + \frac{1}{2} H^{\a\ad} (\nabla_\a {\bar \nabla}_\ad \G - {\bar \nabla}_\ad \nabla_\a {\bar \G} ) 
+ \bar \G \G + \G^2 +{\bar \G}^2 \Big\} ~,
\label{non-action}
\eea
with $\G$ the complex linear compensator obeying the constraint
\bea
({\bar \nabla}^2 -4R ) \G =0~.
\eea
This action is invariant under the gauge transformations 
\begin{align}
\delta H_{\alpha \dalpha} = \nabla_\alpha {\bar L}_\dalpha - \bar \nabla_\dalpha L_\alpha ~, 
\qquad \d \bar \G = -\frac{1}{4} { \nabla}^\a {\bar \nabla}^2 { L}_\a~,
\end{align}
and is dual to the linearized theory  (\ref{eq_N1OldMin}).
Unlike the situation in Minkowski superspace, where an infinite family of non-minimal 
supergravity actions exists, with the corresponding supercurrents being given by eq.  (\ref{conservation-non}),
the theory (\ref{non-action}) proves to be the only dual formulation 
of  the old minimal model  (\ref{eq_N1OldMin}).

Our consideration shows that the structure of the $\cN=1$ AdS supercurrent multiplets  is more 
restrictive than in the super-Poincar\'e case. In what follows, we will study a consistent 
$\cN=2$ AdS supercurrent.

This paper is organized as follows. In section 2, we consider $\N=2$
supergravity with a cosmological constant and construct its solution,
which corresponds to an AdS geometry.
In section 3, we derive the linearized AdS supergravity action.
Its $\N=1$ reduction is the topic of section 4. Both of these
results are generalizations of our previous work \cite{n2_sugra}.
We address some general issues regarding $\N=2$ supercurrents in section 5 and
postulate the general form of the supercurrent for $\N=2$ supergravity + matter
theories coupled to vector and tensor compensators.
There are five technical appendices. Appendices A and B review briefly the
geometry of $\N=1$ and $\N=2$ superspace with structure group
$\rm SL(2,\mathbb C) \times \rm U(\N)_R$. 
Appendix C reviews the improved $\cN=2$ tensor multiplet in curved projective superspace.
Appendix D contains further technical details
of the derivation of the linearized $\N=2$ supergravity action. Similarly, Appendix E
provides further details about the $\N=1$ reduction procedure.

\section{$\N=2$ supergravity with a cosmological term}

In this section we discuss the $\cN=2$ supergravity formulation of \cite{deWPV} 
using the superspace approach of \cite{n2_sugra_tensor}.
This supergravity formulation makes use of two compensators:
 the vector multiplet \cite{GSW} and the tensor multiplet \cite{Wess}.

\subsection{Conformal compensators} 
The vector multiplet can be described in curved superspace by its
covariantly chiral field strength $\cW$  subject to the Bianchi identity\footnote{Such a superfield 
is often called reduced chiral.} \cite{GSW,Howe}
\bea
\cDB^\ad_i \cW= 0~, \qquad
\S^{ij} := \frac{1}{ 4} \Big(\cD^{\a(i}\cD_\a^{j)}+4S^{ij}\Big) \cW&=&\frac{1}{ 4} 
\Big(\cDB_\ad{}^{(i}\cDB^{j) \ad}+4\bar{S}^{ij}\Big)\bar{\cW} ~,
\label{1.1}
\eea
where $S^{ij} $ and ${\bar S}^{ij} $  are special dimension-1 components of 
the torsion.\footnote{Our
 curved-superspace conventions follow  Ref. \cite{KLRT-M2}. In particular,  
we use the superspace geometry of $\cN=2$ conformal supergravity
introduced in \cite{Howe}  (see also \cite{Muller})  in which 
the  structure group is ${\rm SL}(2,{\mathbb C})\times {\rm U}(2)$. 
The relevant information about Howe's formulation is collected in Appendix \ref{Appendix A}.
In what follows, we will use the notation:
$\cD^{ij}:= \cD^{\a(i}\cD_\a^{j)}$ and ${\bar \cD}^{ij} := \cDB_\ad{}^{(i}\cDB^{j) \ad}$.
It should be noted that Howe's realization of $\cN=2$ conformal supergravity 
\cite{Howe} is a simple extension of Grimm's formulation \cite{Grimm} with the structure group  
${\rm SL}(2,{\mathbb C})\times {\rm SU}(2)$. The precise relationship between these two formulations 
is spelled out  in \cite{KLRT-M2}.}
The superfield $\Sigma^{ij}$ is real, ${\bar \S}_{ij} :=(\S^{ij})^* =\ve_{ik}\ve_{jl}\S^{kl}$, and obeys 
the constraints
\begin{align}
\cD^{(i}_\a \Sigma^{jk)} =  {\bar \cD}^{(i}_\ad \Sigma^{jk)} = 0~.
\end{align}
These constraints are characteristic  of the $\cN=2$ linear multiplet 
\cite{Sohnius:1978fw,BS,SSW}.
In  off-shell formulations of  $\cN=2$ supergravity, one of the compensators is usually a vector
multiplet such that its  field strength $\cW$ is nowhere vanishing, $\cW \neq 0$.

There are several ways to realize $\cW$ as a gauge invariant field
strength.
One possibility, which we will use in what follows, is to introduce
the curved-superspace extension \cite{n2_sugra_tensor} of Mezincescu's
prepotential \cite{Mezincescu} (see also \cite{HST}),  $\cV_{ij}=\cV_{ji}$,
which is an unconstrained real SU(2) triplet.\footnote{Another realization for $\cW$, 
in terms of the  weight-zero tropical prepotential in projective superspace, 
is briefly mentioned in Appendix C. A more extensive discussion of this realization
can be found, e.g.,  in Appendix E of \cite{n2_sugra_tensor} where its relation
to the Mezincescu prepotential is derived.}
The expression for $\cW$ in terms of $\cV_{ij}$ \cite{n2_sugra_tensor} is
\begin{align}\label{eq_WMezin}
\cW = \frac{1}{4}\bar\Delta \Big({\cD}^{ij} + 4 S^{ij}\Big) \cV_{ij}~,
\end{align}
where $\bar\Delta$ is the chiral projection operator \eqref{chiral-pr}. 
Note that $\cV_{ij}$ is defined only up to gauge transformations of the form
\begin{align}
\delta \cV^{ij} &= \cD^{\alpha}{}_k \Lambda_\alpha{}^{kij}
     + \bar\cD_{\dalpha}{}_k \bar\Lambda^\dalpha{}^{kij}, \qquad
     \Lambda_\alpha{}^{kij} = \Lambda_\alpha{}^{(kij)}~,
     \qquad \bar\Lambda^\dalpha{}_{kij} := ( \Lambda_\alpha{}^{kij} )^*~,
\label{pre-gauge1}
\end{align}
with the gauge parameter $ \Lambda_\alpha{}^{kij} $ being completely arbitrary modulo 
the algebraic condition given.

The tensor  (or {\it linear}) multiplet can be described in curved superspace by
its gauge invariant field strength $\cG^{ij}$  which is defined to be a  real ${\rm SU}(2)$ triplet (that is, 
$\cG^{ij}=\cG^{ji}$ and ${\bar \cG}_{ij}:=(\cG^{ij})^* = \cG_{ij}$)
subject to the covariant constraints  \cite{BS,SSW}
\bea
\cD^{(i}_\a \cG^{jk)} =  {\bar \cD}^{(i}_\ad \cG^{jk)} = 0~.
\label{2.5}
\eea
These constraints are solved in terms of a covariantly chiral
prepotential $\Psi$ \cite{HST,GS82,Siegel83,Muller86} as follows:
\begin{align}
\label{eq_Gprepotential}
\cG^{ij} = \frac{1}{4}\Big( \cD^{ij} +4{S}^{ij}\Big) \Psi
+\frac{1}{4}\Big( \cDB^{ij} +4\bar{S}^{ij}\Big){\bar \Psi}~, \qquad
{\bar \cD}^i_\ad \J=0~.
\end{align}
The prepotential is defined up to  gauge transformations of the form
\bea\label{eq_PsiGauge}
\d \J = {\rm i} \,\L ~, \qquad 
\Big(\cD^{ij}+4S^{ij}\Big) \L&=& 
\Big(\cDB^{ij} + 4\bar{S}^{ij}\Big)\bar{\L} ~,
\eea
with $\Lambda$ an arbitrary  reduced chiral superfield.

If the tensor multiplet is chosen as one of the two supergravity compensators, 
then the scalar 
\bea
\cG := \sqrt{\frac{1}{2} \cG^{ij} \cG_{ij}}
\eea
must be nowhere vanishing, $\cG \neq 0$.

\subsection{Dynamics in supergravity}

In accordance with the analysis given  in \cite{n2_sugra_tensor},  
the gauge-invariant supergravity action can be written as 
\begin{subequations}\label{eq_N2NonlinearAction}
\bea
S
&=& \frac{1}{ \k^2} \int \rd^4 x \,{\rm d}^4\q \, \cE \, \Big\{
\J {\mathbb W} - \frac{1}{4} \cW^2 -\x \J \cW \Big\}          +{\rm c.c.}   \label{2.8a} \\
 &=& \frac{1}{ \k^2} \int \rd^4 x \,{\rm d}^4\q \, \cE \, \Big\{
\J {\mathbb W} - \frac{1}{4} \cW^2 \Big\} +{\rm c.c.}
- \frac{\x}{ \k^2} \int \rd^4 x \,{\rm d}^4\q \, \rd^4\bar\theta\,
     E \, \cG^{ij}\cV_{ij}~,~~~~~~~
\label{2.8b} 
\eea
\end{subequations}
where 
\bea
\mathbb W := -\frac{\cG}{8} (\cDB_{ij} + 4 \bar S_{ij}) \left(\frac{\cG^{ij}}{\cG^2} \right) 
\eea
is a composite reduced chiral superfield \cite{deWPV,n2_sugra_tensor}
(that is $\mathbb W$  obeys the same  conditions  (\ref{1.1})  as the field strength $\cW$).
Here $\k$ is the gravitational constant, and $\x$ the cosmological constant.  
In what follows, we will choose $\k =1$.
The first representation for the action, eq. (\ref{2.8a}), involves the integration over the chiral subspace, 
with $\cE$ the chiral density. In the second form, eq. (\ref{2.8b}), the cosmological term 
is given as an integral over the full superspace, with $E^{-1}= {\rm Ber}(E_A{}^M)$.
In Appendix C, we give an alternative expression for the supergravity action.

The equations of motion associated with the above action were derived 
in  \cite{n2_sugra_tensor}.  They are
\begin{subequations}\label{2.11} 
\bea
\cG-\cW \bar \cW&=& 0~, \label{2.11a}  \\
\S^{ij} +\x \cG^{ij} &=&0 ~, \label{2.11b} \\
\mathbb W  -\x \cW &=& 0~. \label{2.11c}
\eea
\end{subequations}
Eq. (\ref{2.11a}) corresponds to the Weyl  multiplet  \cite{deWvHVP,BdeRdeW}, 
i.e. the multiplet of conformal supergravity.
As shown in detail in \cite{KT},  
modulo purely gauge degrees of freedom, 
the Weyl  multiplet 
can be described by a real scalar superfield $\cH$ which we call the gravitation superfield.\footnote{The
appearance of $\cH$ at the linearized supergravity level was revealed in 
\cite{HST,RT,Siegel-curved}.}
The remaining equations (\ref{2.11b}) and (\ref{2.11c}) correspond to the vector
and the tensor compensators, respectively.

\subsection{Solution to the equations of motion: AdS geometry}\label{AdsSolution}
A simple solution to the supergravity equations (\ref{2.11a})--(\ref{2.11c}) can be obtained
in the case that the supersymmetric Weyl tensor $W_{\a \b}$ is zero, 
\be
W_{\a\b}=0~,
\ee
which corresponds to a conformally flat superspace.
Such a solution is easy to derive explicitly by using the super-Weyl gauge
freedom to fix $\cW \bar \cW = \cG$ to a positive constant, which we denote $g$.
This will imply that coordinate dependence of $\cW$ lies only in its phase; similarly,
$\cG^{ij}$ will vary only in the direction it points in isovector space.
Both of these residual degrees of freedom can be fixed by using
almost\footnote{The $\rm SU(2)_R$ gauge freedom is only partially fixed
if we take $\cG^{ij}$ to a constant. A residual $\rm U(1) = \rm SO(2)$
remains, which corresponds to rotations about the axis of $\cG^{ij}$.}
all of the local $\rm U(2)_R$ invariance, but we wish to
examine the consequences of leaving the $\rm U(2)_R$ gauge freedom
unfixed for now.

We consider first the consequences of super-Weyl gauging  $\cW \bar \cW$ to be constant, 
for some reduced chiral superfield $\cW$ such that $\cW \neq 0$. It is immediately apparent
that $\cW$ must itself be 
annihilated by
all the spinor covariant derivatives, 
$\cD_\alpha{}^i \cW = 0$,
by applying the covariant derivative to 
$\cW \bar \cW = g={\rm const}$.
It follows from the algebra of covariant derivatives, eq. (\ref{a-c2}),  that \cite{Howe, KLRT-M2}
\begin{align}
G_{\alpha \dbeta}^{ij} = 0 ~,\label{W=1-a}
\end{align}
as well as 
$\cD_{\alpha \dalpha} \cW = -2\ri \,G_{\alpha \dalpha} \cW.$
In conjunction with the properties $\cD_\alpha{}^i \cW = {\bar \cD}^\ad_i \cW=0$,
the latter relation immediately allows us to solve for the $\rm U(1)_R$
connection\footnote{The contribution of $G_{\alpha \dalpha}$ to 
the connection $\Phi_{\alpha \dalpha}$ is the result of a conventional
constraint. One may redefine the connection to
eliminate this term if one desires.}
\begin{align}
\F_A = -\frac{\ri}{4} E_A  \log (\cW / \bar\cW) +\d_A{}^b \, G_b~,
\label{2.14}
\end{align}
with $E_A =E_A{}^M \pa_M$ the  vielbein, see Appendix A. 
It also follows from the reduced chirality condition (\ref{1.1}) that 
\bea\label{W=1-b}
\cW S^{ij} =  \bar \cW \bar{S}^{ij}~.
\eea

Using the local $\rm U(1)_R$ symmetry allows us to 
gauge away the phase of $\cW$,
that is to impose the gauge condition $\cW = w ={\rm const} $, and then the $\rm U(1)_R$
connection simplifies dramatically, as follows from (\ref{2.14}).

Now we turn to the tensor multiplet and consider the consequences of
enforcing the super-Weyl condition  $\cG = g ={\rm const}$. Using the constraints of
the $\N=2$ tensor multiplet \eqref{2.5}, one may show \cite{n2_sugra_tensor}
that $\cG^{ij}$ is annihilated by the spinor covariant derivatives,
$\cD_\alpha{}^k \cG^{ij} = \bar\cD_\dalpha{}^k \cG^{ij} = 0$. From
\eqref{a-c1} and \eqref{a-c2}, one may show that
$\cD_{\alpha \dalpha} \cG^{ij} = 4 G_{\alpha \dalpha}^{k (i} \cG_k^{j)}$
as well as
\begin{align}
G_{\alpha \dalpha} = Y_{\alpha \beta} = 0~, \qquad
S^{ij} \propto \cG^{ij}~.
\end{align}
The vanishing of the spinor derivatives of $\cG^{ij}$ is a powerful condition;
it implies a set of conditions on the $\rm SU(2)_R$ connection\footnote{The contribution
of $G_{\alpha \dalpha}^{ij}$ to $\Phi_{\alpha \dalpha}^{ij}$
is the consequence of another conventional constraint.}
\begin{align}
\Phi_{A}^{kl} \left(\delta^{ij}_{kl} - \frac{1}{2 g^2} \cG_{kl} \cG^{ij} \right)
     &= \frac{1}{2 g^2} 
     \cG^{ik} E_A \cG_k{}^j
     - 2 \d_A{}^bG_{b}^{kl}\left(\delta^{ij}_{kl} - \frac{1}{2 g^2} \cG_{kl} \cG^{ij} \right)~.
\label{2.17}
\end{align}
Note that this determines the isospin connection only along directions perpendicular
to $\cG^{ij}$; the parallel component gauges rotations about the axis of $\cG^{ij}$
and is completely undetermined. 

Using the local $\rm SU(2)_R$ symmetry allows us to turn
the covariantly constant $\cG^{ij}$ 
into a truly constant isovector, $\cG^{ij} = g^{ij}={\rm const}$. 
When such a gauge is chosen, 
the first term on the right of (\ref{2.17}) drops out.

It remains to enforce the equation of motion $\cG = \cW \bar \cW$ which
allows both sets of the above conditions to be applied simultaneously.
This implies, in particular, that the only torsion superfield
is $S^{ij}$, which along with $\cW$ and $\cG^{ij}$ are all covariantly constant.
From the other two equations of motion \eqref{2.11b} and \eqref{2.11c},
one may read off the solution
\begin{align}
S^{ij} = -\xi \cG^{ij} / \cW~, \qquad
\bar S^{ij} = -\xi \cG^{ij} / \bar \cW~.
\end{align}
The superspace background we have found is maximally symmetric
with the covariant derivatives obeying the algebra \cite{KLRT-M1,KT-M-ads}:
\begin{subequations}
\begin{gather}
\{\cD_\alpha{}^i, \cD_\beta{}^j\} = 4 S^{ij} M_{\alpha \beta}
     + 2 \eps_{\alpha \beta} \eps^{ij} S^{kl} J_{kl}~, \qquad
\{\cD_\alpha{}^i, \bar\cD_\dalpha{}_j\} = -2\ri \delta^i_j \cD_{\alpha \dalpha}~, \\
[\cD_\beta{}^i, \cD_{\alpha \dalpha}] = -\ri \eps_{\beta \alpha} S^{ij} \bar\cD_{\dalpha j}~, \qquad
[\cD_a, \cD_b] = -S^2 M_{ab}~,
\end{gather}
\end{subequations}
where we have denoted $S^2 :=  \frac{1}{2} S^{ij} \bar S_{ij} = \xi^2 g$. From the explicit
form of the Riemann tensor, it is clear that the space-time geometry is AdS, with the
curvature scale determined by the magnitude of the constant $\xi^2 g$.

\section{Linearized AdS supergravity action}

Our goal is to linearize the supergravity action (\ref{2.8a}) or (\ref{2.8b}) around AdS superspace
which has been shown to be an exact solution of the  supergravity equations of motion 
 (\ref{2.11a})--(\ref{2.11c}). 
We represent the compensators in the form
\begin{subequations}
\begin{align}
\cG^{ij} &\rightarrow \cG^{ij} + \qG^{ij}~, \\
\cW &\rightarrow \cW + \qW~,
\end{align}
\end{subequations}
where $\cG^{ij}$ and $\cW$ on the right hand side 
correspond to the covariantly constant background compensators,
while  $\qG^{ij}$ and $\qW$ are arbitrary deformations obeying their
respective Bianchi identities. They may be represented in terms
of linearized prepotentials via
\begin{subequations}
\begin{align}
\qG^{ij} &= \frac{1}{4} (\cD^{ij} + 4 S^{ij}) \qPsi + \frac{1}{4} (\bar\cD^{ij} + 4 \bar S^{ij}) \bar\qPsi ~,
\qquad {\bar \cD}^\ad_i {\bm \J} =0~, \label{3.3} \\
\qW &= \frac{1}{4} \bar\Delta (\cD^{ij} + 4 S^{ij}) \qV_{ij}~, \qquad
 {\bf V}_{ij} = {\bf V}_{ji} = ({\bf V}^{ij})^* ~.  \label{3.4}
\end{align}
\end{subequations}
We also introduce $\qH$ for the linearized gravitational superfield.\footnote{The background 
covariant derivatives depend on
some background prepotentials. However, the explicit form of such a dependence is not essential 
for our purposes.}

\subsection{Linearized supergravity gauge transformations}
We postulate the linearized supergravity gauge transformations:
\begin{subequations}\label{eq_AdsGauge}
\begin{align}
\delta \qPsi &= 4 \bar\Delta(\bar\qOmega^{ij} \cG_{ij}) ~, \label{3.5a} \\
\delta \qV^{ij} &= -4\qOmega^{ij} \bar \cW - 4 \bar \qOmega^{ij} \cW ~,  \label{3.5b}\\
\delta \qH &= (\cD^{ij} + 4 S^{ij}) \qOmega_{ij} + 
     (\bar\cD^{ij} + 4 \bar S^{ij}) \bar \qOmega_{ij}~, \label{3.5c}
\end{align}
\end{subequations}
as  natural generalizations of those given in \cite{n2_sugra}.
The rule for $\qV^{ij}$ is exactly as in the Minkowski background,
while that for $\qPsi$ is the only possible generalization when
we take into account that $\delta\qPsi$ \emph{must} be covariantly chiral.
Note that the formulae (\ref{3.5a}) and (\ref{3.5b}) 
are background super-Weyl covariant if $\qOmega_{ij}$
possesses weight $-3$ with $\qPsi$ and $\qV^{ij}$ having weights
$1$ and $-2$, respectively.

There remains some arbitrariness in the choice of $\delta \qH$,
in particular the choice of the numerical factor in front of $S^{ij}$.
The particular choice made in (\ref{3.5c})
is the one respecting background super-Weyl
covariance when $\qH$ is assumed to transform with weight $-2$.
That the variations should be background super-Weyl covariant is reasonable
since the original theory is super-Weyl invariant, but
we can marshal some additional evidence for this. For instance, the supergrvity 
equations of motion (\ref{2.11a})--(\ref{2.11c})  arise from the first order action
\begin{align}
S^{(1)} &= \int \rd^4x\, \rd^4\theta\, \cE \,\qPsi (\mathbb W - \xi \cW)
     + \int \rd^4x\, \rd^4\bar\theta\, \bar{\cE} \,\bar\qPsi (\bar{\mathbb W} - \xi \bar\cW)
     \eol & \quad
     + \int \rd^4x\, \rd^4\theta\, \rd^4\btheta\, E\, \Bigg\{
     \qH \Big( \cG - \cW \bar \cW\Big)
     - \qV^{ij} \Big(\Sigma_{ij} + \xi \cG_{ij}\Big)\Bigg\}
\end{align}
which is gauge invariant \emph{precisely} for the choice \eqref{eq_AdsGauge}.
More generally, in any superconformal theory $\qH$ couples to a
supercurrent $\cJ$ which is itself super-Weyl covariant with weight $+2$;
in the above example $\cJ = \cG - \cW \bar \cW$. For this linearized
coupling to be sensible, $\qH$ must share this covariance property
and have weight $-2$.

${}$From eqs. (\ref{3.5a}) and (\ref{3.5b}) we read off the supergravity gauge transformations of 
the linearized field strengths (\ref{3.3}) and (\ref{3.4}):
\begin{subequations}
\bea
\d \qG^{ij} &=&  (\cD^{ij} + 4 S^{ij}) \bar\Delta(\bar\qOmega^{ij} \cG_{ij}) 
+  (\bar\cD^{ij} + 4 \bar S^{ij}) \Delta(\qOmega^{ij} \cG_{ij})~, \\
\d \qW &=& - \bar\Delta (\cD^{ij} + 4 S^{ij}) (\qOmega^{ij} \bar \cW +  \bar \qOmega^{ij} \cW)~.
\label{var-boldW}
\eea
\end{subequations}

It should be emphasized that, in this subsection,  
no assumption has been made about the background fields chosen.
The linearized  supergravity gauge transformations (\ref{3.5a})--(\ref{3.5c}) hold for an arbitrary 
supergravity background generated by some covariant derivatives $\cD_A$  and  compensators  
$\cG^{ij}$, $\cW$ and  $\bar \cW$.

\subsection{Linearized supergravity action}
We are now prepared to derive the linearized supergravity action around the AdS background 
as described in subsection \ref{AdsSolution}.
It can be uniquely constructed
by including all  the terms quadratic in the compensators $\qPsi$, $\qV^{ij}$
and the gravitational superfield  $\qH$ with the coefficients chosen in such a way as 
 to render a gauge invariant result. 
A several-day calculation\footnote{Its details are collected in Appendix D.} 
leads to the  linearized AdS supergravity action:
\begin{align}\label{eq_N2action}
S^{(2)} &= \int \rd^4x\, \rd^4\theta\,\cE\, \left(-\frac{1}{4} \qW \qW
          + \qPsi \hat{\mathbb W} - \xi \qPsi \qW \right) + \textrm{c.c.}
     \eol & \quad
     + \int \rd^4x\, \rd^4\theta\, \rd^4\btheta\, E\, \Bigg\{
     - \bar \cW \qW \qH - \cW \bar \qW \qH
     + \frac{1}{2g} \cG_{ij} \qG^{ij} \qH 
     \eol & \quad \quad
     -\frac{1}{2} \bar \cW^2 \qH \bar\Delta \qH
     - \frac{1}{2} \cW^2 \qH \Delta \qH
     - \frac{g}{8} \qH S^{ij} \BCD_{ij} \qH
     - \frac{g}{8} \qH \bar S^{ij} \cD_{ij} \qH
     \eol & \quad \quad
     -\frac{1}{64g} \cG_{ij} \cG^{kl} \qH \cD^{ij} \BCD_{kl} \qH
     -\frac{g}{32} \qH \cD^{ij} \BCD_{ij} \qH + \frac{g}{2} \qH \Box \qH\Bigg\}~,
\end{align}
where $\Box = \cD^a \cD_a$ and 
\begin{align}
\hat{\mathbb W} = - \frac{1}{24 g} \BCD_{ij} \qG^{ij}~.
\end{align}
One can check that $\hat{\mathbb W} $
is a reduced chiral superfield, 
\bea
\cDB^\ad_i \hat{\mathbb W}= 0~, \qquad
(\cD^{ij} +4S^{ij} ) \hat{\mathbb W}
&=& (\cDB^{ij} +4\bar{S}^{ij}) \hat{\bar {\mathbb W} }~.
\eea
The linearized supergravity action (\ref{eq_N2action}) is one of the main results of 
our work. In the rigid supersymmetric limit it reduces to the action constructed in \cite{n2_sugra}.

It is natural to choose units so that
\begin{align}
\cW \bar \cW = \cG = g = 1
\end{align}
which implies that $\cW$ is a pure phase superfield and
$\cG^{ij}$ is a unit isovector superfield. We will make this assumption
from section \ref{section4}  on.

It is worth emphasizing that the background compensators $\cW$ and $\cG^{ij}$
in the above action are covariantly constant.\footnote{This form for the $\N=2$
action with covariantly constant compensators is closely related to the
$\N=1$ procedure advocated in \cite{Butter:2009wy}.}
In particular, they can be made \emph{truly} constant by choosing a specific
$\rm U(2)_R$ gauge. In our treatment of the Minkowski case \cite{n2_sugra},
we implicitly took this gauge, with $\cW = w$ and $\cG^{ij} = g^{ij}$.
We will frequently find it useful to refer to this gauge; it should be
clear from context (i.e. the appearance of $w$ and $g^{ij}$ in formulae)
when we are using it.

\subsection{Dual versions of the supergravity action}
Before moving on to the $\N=1$ reduction, we will briefly discuss
a duality which may be applied to the supergravity action,
both in its nonlinear \eqref{eq_N2NonlinearAction} and
linearized \eqref{eq_N2action} forms.

We review first the nonlinear version. We begin by writing the
action \eqref{eq_N2NonlinearAction} with an additional complex
parameter $\alpha$ with unit real part,
\begin{align}
S = \int \rd^4 x \,{\rm d}^4\q \, \cE \, \Big\{
\J {\mathbb W} - \frac{\alpha}{4} \cW^2 -\x \J \cW \Big\} +{\rm c.c.}~, \qquad \alpha + \bar \alpha = 2~.
\end{align}
The imaginary part of $\alpha$ is physically irrelevant at first.
Next we relax $\cW$ to a general chiral superfield and
enforce its reduced chirality using a Lagrange multiplier $\cW_D$,
which is a reduced chiral superfield:
\begin{align}
S = \int \rd^4 x \,{\rm d}^4\q \, \cE \, \Big\{
     \J {\mathbb W} - \frac{\alpha}{4} \cW^2 -\x \J \cW
      + \ri \cW\cW_D
     \Big\} +{\rm c.c.}
\end{align}
Performing the duality, we find
\begin{align}
S = \int \rd^4 x \,{\rm d}^4\q \, \cE \, \Big\{
     \Psi \mathbb{W} + \frac{\xi^2}{\alpha} \left(\Psi - \frac{\ri}{\xi} \cW_D\right)^2
     \Big\} + \textrm{c.c.}
\end{align}
For nonzero $\xi$ it is clear that $\cW_D$ is a Stueckelberg field for $\Psi$.
We can absorb it into $\Psi$, by applying a finite gauge transformation (\ref{eq_PsiGauge}),
 and then we end up with a massive tensor compensator
\begin{align}
S = \int \rd^4 x \,{\rm d}^4\q \, \cE \, \Big\{
     \Psi \mathbb{W} + \frac{\xi^2}{\alpha} \Psi^2
     \Big\} + \textrm{c.c.}
\end{align}
If we parametrize $\alpha = 1 - \ri e / \mu$ with $e^2 + \mu^2 = 4 \xi^2$, the
mass-like term for $\Psi$ becomes $\xi^2 / \alpha = \mu (\mu + \ri e) / 4$,
which can be interpreted as a combination of magnetic and electric
contributions 
which are associated with the two possible 
mass terms $B \wedge {}^* B$ 
and $B \wedge B$ for the component two-form $B$ (see, e.g, \cite{DF} for a pedagogical discussion).
As discussed in \cite{n2_sugra_tensor},
this is a formulation for $\N=2$ supergravity with a cosmological constant and
a single chiral compensator $\Psi$.\footnote{The vector multiplet has been eaten up by the tensor multiplet 
which is now massive. The vector compensator acts as a Stueckelberg field to give mass to the tensor multiplet.
This is an example of the phenomenon observed originally in \cite{LM} and studied in detail 
in \cite{Dall'Agata:2003yr,DSV,DF,LS,Theis:2004pa,Kuzenko:2004tn,K-dual08,Gauntlett:2009zw}.}

We can perform the same duality at the linearized level. The result is
\begin{align}\label{eq_N2Dualaction}
S^{(2)} &= \int \rd^4x\, \rd^4\theta\,\cE\, \left(
     \qPsi \hat{\mathbb{W}}
     + \frac{1}{\alpha} \left(\xi \qPsi - \ri \qW_D\right)^2
     \right) + \textrm{c.c.}
     \eol & \quad
     + \int \rd^4x\, \rd^4\theta\, \rd^4\btheta\, E\, \Bigg\{
     \frac{2\bar \cW}{\alpha} \left(\xi \qPsi - \ri \qW_D\right) \qH
     + \frac{2 \cW}{\bar\alpha} \left(\xi \bar\qPsi + \ri \bar \qW_D\right) \qH
     + \frac{1}{2g} \cG_{ij} \qG^{ij} \qH 
     \eol & \quad \quad
     + \frac{\bar \cW^2}{2} \frac{\bar\alpha}{\alpha} \qH \bar\Delta \qH
     + \frac{\cW^2}{2} \frac{\alpha}{\bar\alpha} \qH \Delta \qH
     - \frac{g}{8} \qH S^{ij} \BCD_{ij} \qH
     - \frac{g}{8} \qH \bar S^{ij} \cD_{ij} \qH
     \eol & \quad \quad
     -\frac{1}{64g} \cG_{ij} \cG^{kl} \qH \cD^{ij} \BCD_{kl} \qH
     -\frac{g}{32} \qH \cD^{ij} \BCD_{ij} \qH + \frac{g}{2} \qH \Box \qH\Bigg\}~.
\end{align}
Under the linearized supergravity gauge transformations, eqs. (\ref{3.5a}) and (\ref{3.5b}), 
 the Stueckelberg field $\qW_D$ transforms as
\begin{align}
\delta \qW_D = -\frac{\ri }{2}  \bar\Delta (\cD^{ij} + 4 S^{ij}) 
(\bar\alpha \qOmega_{ij} \bar \cW
- \a  \bar\qOmega_{ij}  \cW )
 ~,
\end{align}
compare with (\ref{var-boldW}).
As before, we may redefine $\qPsi$ to ``eat'' $\qW_D$.
In the gauge $\qW_D =0$ the supergravity gauge transformation of $\bm \J$, eq. (\ref{3.5a}),  turns into 
\begin{align}
\delta \Psi = 2 \bar\alpha \bar\Delta (\cG^{ij} \bar\qOmega_{ij} + \cG^{ij} \qOmega_{ij})
     + \frac{1}{2\xi} \bar\Delta \cD^{ij} (\alpha \cW \qOmega_{ij} - \bar\alpha \bar\cW \qOmega_{ij})~.
\end{align}
Of course, the transformation of ${\qG}^{ij}$ does not change.

It is also possible to perform a duality to a massive vector multiplet
(i.e. with a polar multiplet acting as a Stueckelberg field). This formulation
lives naturally in projective superspace, so we won't attempt to describe
it here. The nonlinear version was discussed in \cite{K-dual08}.

\section{$\N=1$ reduction}\label{section4}

In our previous paper \cite{n2_sugra}, we considered the
$\N=1$ reduction of the $\N=2$ supergravity without a cosmological
constant (i.e. $\xi=0$). This was an easy procedure  since the background was
Minkowski and the reduction was quite straightforward.
The situation in AdS is markedly different and there are in
principle two ways we might proceed. One is to treat AdS
as a special case of a general curved space and construct
explicitly the $\N=1$ local superspace reduction of a
generic $\N=2$ action. Needless to say this would be a
difficult task, even given the effort already expended 
on $\N=1$ reductions in superspace \cite{Gates:1984wc,
Labastida:1984qa, Labastida:1984us, Labastida:1986md}.

The other approach is to exploit the fact that AdS is a
conformally flat geometry. That is, we can always go to
a super-Weyl gauge where we have flat superspace geometry and
a non-vanishing compensator field.

This is most easily illustrated by the $\N=0$ case.
One can construct Einstein gravity by taking conformal
gravity in the presence of a real scalar field $\phi$ of
unit conformal dimension acting as a conformal
compensator.\footnote{See for example the discussions in
\cite{Fradkin:1978yw, Dirac:1973gk}, but the idea goes back to Weyl.}
(The metric  $g_{mn}$ has conformal dimension $-2$ in this picture.)
One may describe a conformally flat geometry using the set of fields
\begin{align}\label{eq_easy_ads}
g_{mn} = \eta_{mn} e^{2 \Omega} , \qquad \phi = 1~.
\end{align}
This is the conventional picture, with the field $\phi$ essentially
playing no role. (It is placed in any given action so that
Weyl invariance is formally maintained.) We call this the
``Einstein frame.'' Alternatively, one may perform a Weyl transformation to
the set of fields
\begin{align}\label{eq_easy_flat}
g_{mn} = \eta_{mn}~, \qquad \phi = e^{\Omega}~.
\end{align}
In this picture, all of the curvature is contained within $\phi$.
This we refer to as the ``flat frame.''

Physically there is no real difference between these two pictures.
The \emph{effective} metric in the theory is $g_{mn} \phi^2$, which
is Weyl invariant. Our $\N=2$ geometry is only a little more complicated.
Instead of a single compensator field $\phi$, we have $\cG^{ij}$
and $\cW$ which compensate not just for super-Weyl transformations
but also $\rm U(2)_R$ transformations.

For AdS geometry, there exists an explicit solution for $\Omega(x)$
in a certain coordinate chart covering only part of the AdS hyperboloid:
\begin{align}
e^\Omega = \left(1 - \frac{1}{4} \mu^2 x^2\right)^{-1}~.
\end{align}
For this case, we will refer to the Einstein frame as the
``AdS frame.'' An analogous construction exists in superspace.
The $\N=2$ geometry we have described up to this point is
in a conformal frame analogous to the first set of equations
\eqref{eq_easy_ads}. We have nontrivial curvature of all types -- torsion,
Lorentz, \emph{and} isospin -- while our compensator fields
$\cW$ and $\cG_{ij}$ are covariantly constant, and gauge equivalent to
constant values $w$ and $g_{ij}$.
Because we know how to do the $\N=1$ reduction in the case of flat
superspace background (i.e. the analogue of \eqref{eq_easy_flat}),
we will exploit our ability to perform super-Weyl transformations
in both $\N=2$ and $\N=1$ geometries to construct the relation
between $\N=2$ and $\N=1$ AdS actions.

In other words, given a set of covariant derivatives $\cD_\alpha{}^i$
in $\N=2$ AdS, we will relate them to a set of covariant
derivatives $\nabla_\alpha$ in $\N=1$ AdS by the chain
\begin{align}
\cD_\alpha{}^i \xleftarrow[\textrm{super-Weyl}]{} D_\alpha{}^i
     \xrightarrow[\N=1 \,\textrm{reduction}]{} D_\alpha{} = D_\alpha{}^\1
     \xrightarrow[\textrm{super-Weyl}]{} \nabla_\alpha
\end{align}
with the $\N=1$ reduction performed in a frame where it is
straightforward. The same chain of transformations can be applied to any objects
in our theory, including actions. Given a Lagrangian $\cL$ appearing
in the $\N=2$ AdS action
\begin{align*}
S = \int \rd^4x\, \rd^4\theta\, \rd^4\btheta\, E\, \cL
\end{align*}
we may convert it to an $\N=1$ AdS Lagrangian $\cL^{(1)}$ with action
\begin{align*}
S = \int \rd^4x\, \rd^2\theta\, \rd^2\btheta\, E\, \cL^{(1)}
\end{align*}
via the procedure
\begin{align}
\cL \xleftarrow[\textrm{super-Weyl}]{} \cL_0
     \xrightarrow[\N=1 \,\textrm{reduction}]{} \cL_0^{(1)} = \frac{1}{16} (D^\2)^2 (\bar D_\2)^2 \cL_0\vert
     \xrightarrow[\textrm{super-Weyl}]{} \cL^{(1)}~.
\end{align}

Now we need only explicitly construct the transformations taking us to
the flat geometry for $\N=2$ and $\N=1$. We will give first the
$\N=2$ solution, then the $\N=1$ solution, and then describe
how to connect them via a simple reduction procedure.

\subsection{From flat $\N=2$ geometry to AdS}
Conformally flat geometry for $\N=2$ superspace was analyzed in
depth in \cite{KT-M-ads} for the case where the structure group
is $\rm SL(2,\mathbb C) \times \rm SU(2)_R$.\footnote{The 
fact that the $\cN=2$ AdS superspace is locally conformally flat
was also discussed in \cite{BILS}.} 
Here we modify that
presentation somewhat for our choice of structure group
$\rm SL(2,\mathbb C) \times \rm U(2)_R$. We have included 
Appendix \ref{Appendix A} to briefly review the details of
that $\N=2$ superspace.

A conformally flat geometry is defined as any geometry related to a
flat geometry by the combination of super-Weyl and $\rm U(2)_R$
transformations.\footnote{Equivalently,
it is an $\N=2$ geometry obeying the constraint $W_{\alpha\beta} = 0$.}
For our purposes, it will be important only to consider the super-Weyl
transformations. These take the flat space  derivatives $D_A =(\pa_a , D^i_\a, {\bar D}^\ad_i )$ to curved space
 ones $\cD_A =(\cD_a , \cD^i_\a, {\bar \cD}^\ad_i )$ by
\begin{subequations}
\begin{align}
\cD_\alpha{}^i &= e^{\cUnew/2} \left(D_\alpha{}^i + 2 D^{\beta i} \cUnew  M_{\beta \alpha}
     - \frac{1}{2} D_\alpha{}^i \cUnew  \mathbb J + 2 D_\alpha{}^j \cUnew  J_j{}^i\right) ~,\\
\bar\cD^\dalpha{}_i &= e^{\cUnew/2} \left(\bar D^\dalpha{}_i -2 \bar D_{\dbeta i} \cUnew \bar M^{\dbeta \dalpha}
     + \frac{1}{2} \bar D^\dalpha{}_i \cUnew \mathbb J - 2 \bar D^\dalpha{}_j \cUnew J^j{}_i\right)~,
\end{align}
\end{subequations}
where $\cU$ is the super-Weyl parameter. The torsion superfields in the curved
space are given by
\begin{subequations}
\begin{align}
S_{ij} &= \frac{1}{4} e^{3 \cUnew} D_{ij} e^{-2 \cUnew} \label{eq_Sdef} ~,\\
Y_{\alpha \beta} &= -\frac{1}{4} e^{-\cUnew} D_{\alpha \beta} e^{2\cUnew}~, \\
G_{\alpha \dalpha} &= -\frac{1}{16} e^{-\cUnew} [D_\alpha{}^k, \bar D_{\dalpha k}] e^{2\cUnew} ~,\\
G_{\alpha \dalpha}{}^{ij} &= \frac{\ri}{4} e^{\cUnew} [D_\alpha{}^{(i}, \bar D_\dalpha{}^{j)}] \cUnew
     \label{eq_Gijdef}~.
\end{align}
\end{subequations}

The maximally symmetric geometry is AdS, which obeys
the additional constraints
\begin{align}\label{eq_AdsConstraints}
Y_{\alpha \beta} = G_{\alpha \dalpha} = G_{\alpha \dalpha}{}^{ij} = 0~,
\end{align}
which in turn imply that $S^{ij}$ is covariantly constant
\begin{align}
\cD_A S^{ij} =0
\label{4.9}
\end{align}
with constant norm
\begin{align}
S^2 = \frac{1}{2} S^{ij} \bar S_{ij} = \textrm{const}~.
\end{align}
In general, $S^{ij}$ is not \emph{actually} constant; however, one can
always make an additional ${\rm U(2)_R}$ transformation to achieve this.

The constraints \eqref{eq_AdsConstraints} impose a number of additional
conditions on the real parameter $\cU$. For example, using \eqref{eq_Gijdef}
the equation $G_{\alpha \dalpha}{}^{ij} = 0$ is solved by
\begin{align}
\cU = \Sigma + \bar\Sigma~, \qquad {\bar \cD}^\ad_i \S =0~,
\end{align}
for an arbitrary chiral scalar $\Sigma$, neutral under the group ${\rm U(1)_R}$.
This immediately simplifies our search for $\cU$. We temporarily may
convert our AdS covariant derivatives $\cD$ with the structure group
$\rm SL(2,\mathbb C) \times \rm U(2)_R$ to AdS covariant derivatives $\gD$ with
structure group $\rm SL(2,\mathbb C) \times \rm SU(2)_R$ via the similarity
transformation\footnote{The same similarity transformation must be applied to all
superfields, including torsion superfields.}
\begin{subequations}
\begin{align}
\gD_\alpha{}^i &= e^{-\frac{1}{2} (\Sigma - \bar\Sigma) \mathbb J}
     \cD_\alpha{}^i e^{\frac{1}{2} (\Sigma - \bar\Sigma) \mathbb J}
     = e^{\bar\Sigma} \left(D_\alpha{}^i + 2 D^{\beta i} \Sigma\, M_{\beta \alpha}
     + 2 D_\alpha{}^j \Sigma \,J_j{}^i\right)~, \\
\bar\gD^\dalpha{}_i &= e^{-\frac{1}{2} (\Sigma - \bar\Sigma) \mathbb J}
     \bar\cD^\dalpha{}_i e^{\frac{1}{2} (\Sigma - \bar\Sigma) \mathbb J}
     = e^{\Sigma} \left(\bar D^\dalpha{}_i
     - 2 \bar D_{\dbeta i} \bar\Sigma \,\bar M^{\dbeta \dalpha}
     - 2 \bar D^\dalpha{}_j \bar\Sigma \,J^j{}_i\right)~.
\end{align}
\end{subequations}
These AdS covariant derivatives were given in \cite{KT-M-ads}. There the requirements
that their torsions $\mathbb Y_{\alpha \beta}$ and $\mathbb G_{\alpha \dalpha}$
vanish were solved by requiring the chiral superfield $\Sigma$ to obey
\begin{align}\label{eq_sigma}
\exp ( -2 \Sigma ) = \left(1 - \frac{1}{4} s^2 y^2 + s^{ij} \theta_{ij}\right)^{-1}~, \qquad
y^m = x^m + i \theta_j \sigma^m \bar\theta^j~,
\end{align}
where $s^{ij}$ is a constant complex\footnote{Our parameter $s^{ij}$ actually
corresponds to the parameter $b^{ij} = q \mathbf{s}^{ij}$
given in eq (4.16) of \cite{KT-M-ads}. There, $\mathbf s^{ij}$ was a real isotriplet and
$q$ was a complex phase, which was subsequently set to unity. We find it useful
to keep the phase unfixed and consider complex $s^{ij}$.} isovector,
$s^2 := \frac{1}{2} s^{ij} \bar s_{ij}$, and $\theta_{ij} := \theta^\mu_i \theta_{\mu j}$.
We may borrow this solution and use it with our original
$\rm SL(2,\mathbb C) \times \rm U(2)_R$ AdS covariant derivatives.

The complex constants $s^{ij}$ are acted on by the global $\rm U(2)_R$.
Using \eqref{eq_Sdef}, one may show that
\begin{align}
S^{ij} = s^{ij} + \mathcal O(\theta)~, \qquad S^2 = s^2~.
\end{align}
which imply that $S^{ij}$ differs from $s^{ij}$ by a $\theta$-dependent $\rm U(2)_R$ rotation.
Thus we are always free to choose the \emph{local} $\rm U(2)_R$ gauge so that
\begin{align}
S^{ij} \xrightarrow[{\rm U(2)_R}]{} s^{ij}~.
\end{align}
The \emph{global} $\rm U(2)_R$ transformations remain unfixed.

In section \ref{AdsSolution}, we solved the equations of motion
by going to the frame where $\cG^{ij}$ and $\cW$ were covariantly constant
and gauge-equivalent to the constant values $g^{ij}$ and $w$. We can now apply
our super-Weyl transform to work out explicit forms for $\cG^{ij}_0$
and $\cW_0$ in the flat frame. Recall that $\cW$ is a pure phase superfield;
this implies that
\begin{align}
1 = \cW\bar \cW  = e^{2\cUnew} \cW_0 \bar \cW_0~.
\end{align}
This equation can be solved by taking
\begin{align}
\cW_0 = w \,e^{-2\Sigma} = w \left(1 - \frac{1}{4} s^2 y^2 + s^{ij} \theta_{ij}\right)^{-1}
\end{align}
for a constant phase $w$.
This yields an explicit solution for $\cW_0$ in the flat frame.
Note that this solution reduces to $w$ when the AdS curvature goes to zero.

The solution for $\cG^{ij}_0$ can be most readily found by applying
the equation of motion \eqref{2.11b}:
\begin{align}
\cG^{ij}_0 = -\frac{1}{4\xi} D^{ij} \cW_0
     = -\frac{w}{4\xi} D^{ij} \left(1 - \frac{1}{4} s^2 y^2 + s^{ij} \theta_{ij}\right)^{-1}~.
\end{align}
Consistency with the remaining equations \eqref{2.11a} and \eqref{2.11c}
requires that
\begin{align}
s^2 = \xi^2~,
\end{align}
which agrees with the physical requirement that the AdS scale be set
by the cosmological constant. Note that this solution for $\cG^{ij}_0$
tends toward constant $g^{ij} = - w s^{ij} / \xi$ as the AdS curvature
tends to zero.

It is worth noting that 
\begin{align}
\cG_0^{ij} = -\frac{e^{-\cUnew}}{\xi} \cW_0 S^{ij}
\end{align}
and so the same
$\rm U(2)_R$ rotation which sends $S^{ij}$ to constant $s^{ij}$ will
send
\begin{align}
\cG^{ij}_0 \xrightarrow[U(2)_R]{} e^{-2\cUnew} g^{ij}~, \qquad
\cW_0 \xrightarrow[U(2)_R]{} e^{-\cUnew} w~.
\end{align}
Therefore the composition of this $U(2)_R$ with the super-Weyl transformation
does indeed take us to an AdS frame where $\cG^{ij}$ and $\cW$ are
actually constant
\begin{subequations}
\begin{align}
\cG^{ij}_0 &\xrightarrow[{\rm U(2)_R}]{} e^{-2\cUnew} g^{ij} \xrightarrow[\textrm{super-Weyl}]{} g^{ij}~, \\
\cW_0 &\xrightarrow[{\rm U(2)_R}]{} e^{-\cUnew} w \xrightarrow[\textrm{super-Weyl}]{} w~.
\end{align}
\end{subequations}
It turns out we will have no need for the explicit form of this
$\rm U(2)_R$ transformation; it is sufficient to know it exists.

\subsection{From flat $\N=1$ geometry to AdS}
We now consider conformally flat $\N=1$ AdS geometry\footnote{See 
\cite{Keck,Zumino,IS}
for early papers  on $\cN=1$ AdS supersymmetry and  superspace.}
with the structure
group $\rm SL(2,\mathbb C) \times \rm U(1)_R$. The details will
be very similar to the standard discussion with structure group
$\rm SL(2,\mathbb C)$. We have included Appendix \ref{Appendix B}
to briefly review the relevant details of $\N=1$ superspace with
the structure group $\rm SL(2,\mathbb C) \times \rm U(1)_R$.

As with $\N=2$, a conformally flat $\N=1$ geometry can be connected
to a flat geometry via an $\N=1$ super-Weyl + ${\rm U(1)_R}$
transformation.\footnote{Equivalently, a conformally flat $\N=1$ geometry
is characterized by the condition that the  torsion superfield
$W_{\alpha \beta \gamma}$ vanishes.} We are concerned only with the
super-Weyl transformation for now, which acts on the covariant
derivatives,
\begin{subequations}
\begin{align}
\nabla_\alpha &= e^{\Unew / 2} \left(D_\alpha + 2 D^{\beta} \Unew \,M_{\beta \alpha}
     - \frac{3}{2} D_\alpha \Unew \,\hat {\mathbb J} \right) ~, \label{4.23a} \\
\bar\nabla^\dalpha &= e^{\Unew/2} \left(\bar D^\dalpha - 2 \bar D_{\dbeta} \Unew \,\bar M^{\dbeta \dalpha}
     + \frac{3}{2} \bar D^\dalpha \Unew \,\hat {\mathbb J} \right)~,
\end{align}
\end{subequations}
where $U$ is an arbitrary real scalar superfield.\footnote{Note that we
have used a different label $\hat{\mathbb J}$
for the $\rm U(1)_R$ generator than in the $\cN=2$ case.
The  action of $\hat{\mathbb J}$
on the $\cN=1$ covariant derivatives is defined by eq. (\ref{B.3}).
This operator may related to the $\cN=2$  $\rm U(1)_R$ generator
as in eq. (\ref{4.40}).}
The torsion superfields are given by
\begin{subequations}
\begin{align}\label{eq_R}
R &= -\frac{1}{4} e^{3\Unew} \bar \nabla^2 e^{-2\Unew}~, \\
G_{\alpha \dalpha} &= [\nabla_\alpha, \bar \nabla_\dalpha] e^{\Unew} ~,\\
X_\alpha &= -\frac{3}{2} (\bar\nabla^2 - 4 R) \nabla_\alpha \Unew~.
\end{align}
\end{subequations}

The maximally symmetric geometry is AdS, which
is characterized by the additional constraints
\begin{align}\label{eq_Ads1Constraints}
G_{\alpha \dalpha} = X_\alpha = 0~.
\end{align}
These in turn imply that $R$ is covariantly constant
with constant norm
\begin{align}
\nabla_\alpha R = \bar\nabla_\dalpha R = 0~ , \qquad R \bar R = \textrm{const}~.
\end{align}
In general $R$ is not \emph{actually} constant, but a
local $\rm U(1)_R$ transformation can always make it so.

The constraints \eqref{eq_Ads1Constraints} impose a number of conditions
on the super-Weyl parameter $U$, which can be solved by
\begin{align}
U = \sigma + \bar\sigma
\end{align}
for chiral $\sigma$ parametrized by a constant complex parameter $\mu$
\begin{align}
\exp (-2 \sigma) = \left(1 - \frac{1}{4} \mu^2 y^2 - \bar\mu \theta^2\right)^{-1}~, \qquad
y^m = x^m + i \theta \sigma^m \bar\theta~.
\end{align}
Using \eqref{eq_R}, one may show that
\begin{align}
R = \mu + \mathcal O(\theta)~, \qquad R \bar R = |\mu|^2~,
\end{align}
which imply that $R$ differs from $\mu$ by a
$\theta$-dependent $\rm U(1)_R$ rotation; thus we are always
free to choose the \emph{local} $\rm U(1)_R$ gauge
\begin{align}
R \xrightarrow[\rm U(1)_R]{} \mu~.
\end{align}
A residual \emph{global} $\rm U(1)_R$ symmetry acts on $\mu$.

It is also possible to understand $\N=1$ AdS geometry in terms
of a compensator. One couples a chiral compensator $\Phi$ with a
cubic coupling $\xi$ to conformal supergravity with the action
\begin{align}
S = -3 \int \rd^4x\, \rd^2\theta\, \rd^2\btheta\, E\, \Phi \bar\Phi
     + \xi \int \rd^4x\, \rd^2\theta\, \cE\, \Phi^3
     + \xi \int \rd^4x\, \rd^2\btheta\, \bar\cE\, \bar\Phi^3.
\end{align}
The equation of motion in the flat frame is
\begin{align}
-\frac{1}{4} \bar D^2 \bar\Phi_0 = \xi \Phi_0^2
\end{align}
and it has the solution $\Phi_0 = \varphi e^\sigma$ for complex phase
$\varphi$ where $\mu = \xi \varphi^3$ \cite{BK}. In the AdS frame,
it follows that $\Phi = \varphi$; in other words
\begin{align}
\Phi_0 \xrightarrow[\textrm{super-Weyl} + U(1)_R]{} \varphi.
\end{align}

\subsection{Procedure for the reduction to $\N=1$}
The solution for $\N=2$ AdS is parametrized by a constant isovector
$s^{ij} = -\xi g^{ij} / w$ which is rotated by the global $\rm U(2)_R$
action. There are two interesting possibilities to choose for $g^{ij}$.
One is
\begin{align}
g^{\1\1} = g^{\2\2} = 0~,
\end{align}
while the other is
\begin{align}
g^{\1\2} = 0~.
\end{align}
Making a specific choice for $g^{ij}$ is equivalent to choosing
which supersymmetry to leave manifest in the $\N=1$ reduction.
In our previous work \cite{n2_sugra} where we considered $\N=2$
Minkowski superspace (i.e. with vanishing cosmological constant, $\xi=0$),
we found that both of these were sensible choices.\footnote{In fact, it was
possible in \cite{n2_sugra} to perform an $\N=1$ reduction for \emph{any} choice of the
parameter $g^{ij}$.} The first of these conditions corresponded to
linearized new minimal supergravity and the second to linearized
old minimal supergravity, each accompanied by a massless
gravitino multiplet, which are dual to each other.

AdS offers much less freedom in this choice. 
As discussed in section 1,
the linearized $\cN=1$ supergravity action exists in AdS only for two cases
of compensator field: a chiral compensator (corresponding to old
minimal supergravity) and a complex linear compensator (corresponding
to the $n=-1$ non-minimal supergravity), which are dual to each other.
Therefore, we expect that only the choice $g^{\1\2} = 0$ will yield an
elegant $\N=1$ reduction.

Taking $g^{\1\2} = 0$, it is convenient to introduce the complex phase
parameter $\gamma$,
\begin{align}
g_{\1\1} = \gamma~, \qquad g_{\2\2} = \bar\gamma~, \qquad \gamma \bar\gamma = 1~,
\end{align}
yielding
\begin{subequations}
\begin{align}
s^{\1\1} = -\xi \bar\gamma \bar w~, \qquad
s^{\2\2} = -\xi \gamma \bar w ~,\\
\bar s_{\1\1} = -\xi \gamma w~, \qquad
\bar s_{\2\2} = -\xi \bar \gamma w ~.
\end{align}
\end{subequations}
The $\N=2$ chiral super-Weyl parameter $\Sigma$ takes the form
\begin{align}
\exp (-2\Sigma) = \left(1 - \frac{1}{4} \xi^2 y^2 - \xi \bar\gamma \bar w \,\theta_{\1\1}
     - \xi \gamma \bar w \,\theta_{\2\2} \right)^{-1}~.
\end{align}
The coefficients of $\theta_{\1\1}$ and $y^2$ have the same relationship
required by the coefficients of $\theta^2$ and $y^2$ in the $\N=1$
parameter $\sigma$, so we may take $\sigma = \Sigma\vert$. This implies
\begin{align}
\bar\mu = \xi \bar\gamma \bar w = -s^{\1\1}~.
\end{align}

We can now explicitly relate the $\N=2$ and $\N=1$ AdS derivatives.
Observe that the flat space derivative identification $D_\alpha{}^\1 = D_\alpha$
and the relation $\cU | =U$ lead to 
\begin{align}
\cD_\alpha{}^\1 \vert &= e^{\Unew/2} \left(D_\alpha + 2 D^{\beta} \Unew \,M_{\beta \alpha}
     - \frac{1}{2} D_\alpha \Unew \,\mathbb J
     + 2 D_\alpha \Unew \,J_\1{}^\1
     + 2 D_\alpha{}^\2 \cUnew\vert \,J_\2{}^\1 \right)~.
\end{align}
Moreover, because $g^{\1\2} = 0$, one can show that $D_\alpha{}^\2 \cU\vert = 0$.
Comparing the expression for $\cD_\alpha{}^\1 \vert$ with that for the $\cN=1$ covariant derivative $\nabla_\a$, 
eq. (\ref{4.23a}), 
we are led to identify the $\N=1$ $\rm U(1)_R$
generator with  a certain diagonal subgroup of $\rm U(2)_R$:
\begin{align}
\hat{\mathbb J} = \frac{1}{3} \mathbb J - \frac{4}{3} J^\1{}_\1~.
\label{4.40}
\end{align}
This leads to the very simple AdS relations:
\begin{align}\label{eq_DReduction}
\cD_\alpha{}^{\ul 1}\vert = \nabla_\alpha~, \qquad
\bar\cD_\dalpha{}_{\ul 1}\vert = \bar\nabla_\dalpha~, \qquad
\cD_{\alpha \dalpha}\vert = \nabla_{\alpha \dalpha} = \frac{\ri}{2} \{\nabla_\alpha, \bar\nabla_\dalpha\}~.
\end{align}
From the AdS algebra in both cases, we find that
\begin{align}
R: = -\bar S_{\1\1}\vert~, \qquad
\bar R := -S^{\1\1}\vert
\end{align}
are covariantly constant. Moreover, one can show that
\begin{align}
S^{\1\2}\vert = 0~.
\end{align}

We can now perform the general reduction to $\N=1$ of any action.
We begin from the AdS frame where $S^{ij}$, $\cG^{ij}$, and $\cW$
are only covariantly constant, and only a super-Weyl transformation
separates us from the flat frame. The generic action
\begin{align}
S = \int \rd^4x\, \rd^4\theta\, \rd^4\btheta \, E\, \cL
\end{align}
can then be super-Weyl transformed to the flat
frame\footnote{Note that $\cL_0 = \cL$ is unchanged since it has super-Weyl weight
zero; this implies that the AdS $E$ is equal to unity.}
\begin{align}
S = \int \rd^4x\, \rd^4\theta\, \rd^4\btheta \, \cL_0~.
\end{align}
The $\N=1$ reduction in the flat geometry is straightforward:
\begin{align}
S = \int \rd^4x\, \rd^2\theta\, \rd^2\btheta \, \cL_0^{(1)}~, \qquad
     \cL_0^{(1)} = \frac{1}{32} \{(D^\2)^2, (\bar D_\2)^2\} \cL_0\vert ~.
\end{align}
Performing an $\N=1$ super-Weyl transformation back to the AdS
geometry gives
\begin{align}
S = \int \rd^4x\, \rd^2\theta\, \rd^2\btheta \, E\, \cL^{(1)}~, \qquad
\cL^{(1)} = e^{2\Unew} \cL^{(1)}_0~.
\end{align}
Discarding $\N=1$ total derivatives, one can show that
\begin{align}
\cL^{(1)} &= \frac{1}{32} \Big((\cD^\2)^2 + 8 S^{\ul {22}}\Big) (\bar \cD_\2)^2 \cL \vert
     + \frac{1}{32} \Big((\bar \cD_\2)^2 + 8 \bar S_{\ul{22}}\Big) (\cD^\2)^2 \cL \vert~.
\end{align}
This formula can now be directly applied to any $\N=2$ full superspace
action.

This procedure may also be applied to chiral actions.
Beginning with
\begin{align}
S = \int \rd^4x\, \rd^4\theta\, \cE\, \cL_c~,
\end{align}
we can first perform a super-Weyl transformation to the flat frame
\begin{align}
S = \int \rd^4x\, \rd^4\theta\, \cL_{c\,0}~, \qquad \cL_{c\,0} = e^{-2\cUnew} \cL_{c}~.
\end{align}
The $\N=1$ reduction in the flat geometry is simple:
\begin{align}
S = \int \rd^4x\, \rd^2\theta\, \, \cL_{c\, 0}^{(1)}~, \qquad
     \cL_{c\,0}^{(1)} = -\frac{1}{4} (\bar D_\2)^2 \cL_0\vert~.
\end{align}
Performing an $\N=1$ super-Weyl transformation back to the AdS
frame gives
\begin{align}
S = \int \rd^4x\, \rd^2\theta\, \rd^2\btheta \, \cE\, \cL_c^{(1)}~, \qquad
\cL_c^{(1)} = e^{3\Unew/2} \cL^{(1)}_0~.
\end{align}
One can show that
\begin{align}
\cL_c^{(1)} &= -\frac{1}{4} \Big((\bar \cD_\2)^2 + 8 \bar S_{\ul{22}}\Big) \cL_c \vert~.
\end{align}

\subsection{Background superfields and the fixing of local ${\rm U(1)_R}$}
In performing the reductions of the actions, we will need to identify
the $\N=1$ reductions of the various background superfields.
As we have mentioned, the $\rm U(2)_R$ gauge symmetry of the $\N=2$ geometry must first
be fixed so that only a super-Weyl transformation separates us from
the flat geometry. In such a gauge, we have covariantly constant background superfields,
which differ from constant values by a $\theta$-dependent $\rm U(2)_R$ transformation,
\begin{align}
\cW = w + \mathcal O(\theta)~, \qquad
\cG^{ij} = g^{ij} + \mathcal O(\theta)~.
\end{align}
The requirement of covariant constancy is quite powerful since in applying the
reduction formulae to the actions, we use the AdS covariant derivatives
and so no derivatives of $\cW$ or $\cG^{ij}$ can be generated.

The background vector multiplet yields then a single $\N=1$ background chiral
superfield $\cW\vert$.\footnote{For any choice other than $g^{\1\2} = 0$, we would
find a non-vanishing background $\N=1$ vector multiplet field strength $\cW_\alpha$
as well. This is another way to understand the simplicity of the $g^{\1\2} = 0$ choice.}
Because of the relation \eqref{eq_DReduction}, one may
show that $\cW\vert$ is \emph{also} covariantly constant with respect to the
$\N=1$ AdS derivatives and is a pure phase superfield. This means we are
free to fix the $\rm U(1)_R$ gauge freedom by imposing the gauge choice
\begin{align}
\cW\vert = w={\rm const}~.
\end{align}
We will make this choice for simplicity.
Because we have gauged a chiral superfield to a constant, the $U(1)_R$
connections must vanish. It follows that $R$ 
is reduced to an actual constant
\begin{align}
R = \mu ={\rm const} ~.
\end{align}

Similarly, the tensor multiplet obeys the conditions
\begin{align}
\cG_{ij}\vert = -\frac{1}{\xi} \cW S_{ij}\vert = -\frac{1}{\xi} \bar \cW \bar S_{ij}\vert
\end{align}
leading to
\begin{align}
\gamma = \cG_{\1 \1}\vert =  \frac{\bar w \mu}{\xi}~, \qquad
\bar \gamma = \cG_{\2 \2}\vert =  \frac{w \bar\mu}{\xi}~,\qquad
\cG_{\1\2}\vert = 0, \qquad
\gamma \bar\gamma = 1
\end{align}
for the complex constant $\gamma$.

For convenience, we collect the formulae relating the various constants
and background superfields in this gauge:
\begin{gather*}
\cW \vert = w~, \qquad
\cG_{\1\1}\vert = \gamma~, \qquad
\cG_{\2\2}\vert = \bar \gamma~,\qquad
\cG_{\1\2}\vert = 0~, \\
S^{\1\1}\vert = -\bar R = -\bar \mu~, \qquad
\bar S_{\1\1}\vert = -R = -\mu ~,\\
S^{\2\2}\vert = - \mu \bar w^2~, \quad
\bar S_{\2\2}\vert = -\bar \mu w^2 ~,\\
S^{\1\2}\vert = \bar S_{\1\2}\vert = 0~, \\
\mu = \xi \gamma w~, \qquad
\bar\mu = \xi \bar\gamma \bar w~, \qquad
w \bar w = \gamma\bar\gamma = 1~.
\end{gather*}
It is convenient to think of $w$ and $\gamma$ as free phases,
which can be chosen by using the \emph{diagonal} part of the global
$\rm U(2)_R$ which is preserved as a symmetry by our reduction
procedure; the last line in the equations above then defines the
constant $R = \mu$.

\subsection{$\N=1$ superfields and supergravity gauge transformations}
We turn next to the identification of $\N=1$ superfields and the
supergravity gauge transformations.
Recall that we explicitly showed in \cite{n2_sugra}  how to identify the $\N=1$
components of the $\N=2$ supergravity multiplet in Wess-Zumino gauge.
This correspondence was constructed in a Minkowski frame, so we may apply
those rules to performing the $\N=1$ reduction in the flat frame of our
current AdS model. We relegate the details of how precisely to do this to
Appendix \ref{N1_reduction_details} and give only the results.

The supergravity gauge transformations \eqref{eq_AdsGauge} allow us to impose the Wess-Zumino gauge
\begin{align}\label{eq_WZgaugeHAdS}
\qH\vert = \cD_\alpha^{\ul 2} \qH\vert = 
     \bar\cD_\dalpha{}_{\ul 2} \qH\vert = (\cD^{\ul 2})^2 \qH\vert = 
     (\bar \cD_{\ul 2})^2 \qH\vert = 0~.
\end{align}
The remaining components of the superfield $\qH$ may be identified as
\begin{subequations}
\begin{align}
H_{\alpha \dalpha} &:= 
     \frac{1}{4} [\cD_\alpha{}^{\ul 2}, \bar \cD_{\dalpha}{}_{\ul 2}] \qH\vert ~,\\
\Psi_{\alpha} &:= 
     \frac{1}{8} (\bar \cD_{\ul 2})^2 \cD_\alpha{}^{\ul 2} \qH\vert  ~,\\
\hat U &:= \frac{1}{16} \cD^\alpha{}^{\ul 2} (\bar \cD_{\ul 2})^2 \cD_\alpha{}^{\ul 2} \qH \vert
     + \frac{1}{12} [\nabla^\alpha, \bar \nabla^\dalpha] H_{\alpha \dalpha}~.
\end{align}
\end{subequations}
Here $H_{\alpha \dalpha}$ is the $\N=1$ gravitational superfield,
$\Psi_\alpha$ is the spinor superfield associated with the second gravitino, 
and $\hat U$ is an auxiliary real scalar superfield, all in the
$\N=1$ AdS frame.\footnote{In \cite{n2_sugra}, we used the $\N=1$
superfield $U$ which differed in its definition from $\hat U$. It turns
out that $\hat U$ has a much simpler super-Weyl transformation
law than $U$. This is sensible since it is the combination $\hat U$,
rather than $U$, which appears in the $\N=1$ reduction of the superconformal $\N=2$
Noether coupling of $\qH$ to its conserved current, which we constructed
in Appendix B of \cite{n2_sugra}.}

It is quite natural to perform the same super-Weyl transformation on the
flat frame $\N=1$ gauge transformations found in \cite{n2_sugra}.
In the AdS frame, they are
\begin{subequations}\label{eq_N1gauge}
\begin{align}
\delta H_{\alpha \dalpha} &= \nabla_\alpha \bar L_\dalpha - \bar \nabla_\dalpha L_\alpha ~,\\
\delta \Psi_\alpha &= \nabla_\alpha \Omega + \Lambda_\alpha ~, 
\qquad {\bar \nabla}_\ad \L_\a =0~,
\\
\delta \hat U &= \r + \bar \r~, \qquad 
{\bar \nabla}_\ad \r =0~,
\end{align}
\end{subequations}
where $\r$ and $\Lambda_\alpha$ are covariantly chiral,
while  $\Omega$ and $L_\alpha$ are unconstrained complex superfields.\footnote{In our previous paper 
\cite{n2_sugra}, the gauge parameter $\r$ was denoted $\hat\Phi$.}

We may apply the same procedure for identifying the $\N=1$ components
of the other $\N=2$ superfields in AdS.
The components of the $\N=2$ vector multiplet $\qW$ consist
of a chiral scalar $\chi$ and the abelian vector field strength $W_\alpha$
given by
\begin{align}
\chi := \qW \vert~, \qquad W_\alpha := \frac{\ri}{2} \cD_\alpha{}^\2 \qW\vert~.
\end{align}
Similarly, the components of the $\N=2$ tensor multiplet $\qG_{ij}$
are given by a chiral scalar $\eta$ and a tensor multiplet $L$,
\begin{align}
\eta := \qG_{\1\1}\vert~, \quad \bar\eta := \qG_{\2\2}\vert~, \quad L = -2\ri \qG_{\1\2}\vert~.
\end{align}
These $\N=1$ superfields naturally transform under the $\N=1$ supergravity
gauge transformations \eqref{eq_N1gauge}:
\begin{subequations}\label{eq_N1gaugeCompensators}
\begin{align}
\delta \chi &= -\frac{w}{12} (\bar \nabla^2 - 4 R) \nabla^\alpha L_{\alpha}
     -  w \r ~,\\
\delta W_{\alpha} &= \frac{\ri}{4} (\bar \nabla^2 - 4 R) \nabla_\alpha \left(
      w \bar\Omega -  \bar w\Omega \right) ~,\\
\delta \eta &=
     - \frac{\gamma}{6} (\bar \nabla^2 - 4 R) \nabla^\alpha L_{\alpha}
     + \gamma \r~, \\
\delta L &=  \ri \gamma \nabla^\alpha \Lambda_{\alpha} 
     - \ri \bar\gamma \bar \nabla_\dalpha \bar\Lambda^\dalpha~.
\end{align}
\end{subequations}
These are natural generalizations of the Minkowski space results found in
\cite{n2_sugra}. It is clear that $\chi$ and $\eta$ transform as
chiral compensators with conformal dimension 1 and 2, respectively, while
they carry opposite charge under the $\rm U(1)$ transformation gauged by the
$\N=1$ auxiliary superfield $\hat U$.\footnote{This $\rm U(1)$ may be understood
as the shadow chiral rotation discussed in \cite{Kuzenko:2007qy}. It corresponds to
a subset of $\rm U(2)_R$ which rotates $\theta_\2$ while leaving $\theta_\1$
invariant.}

\subsection{The $\N=1$ action}
We now give the action corresponding to the $\N=1$ reduction of \eqref{eq_N2action}.
It consists of four pieces, $S = S_{WG} + S_{WH} + S_{GH} + S_{HH}$.
The terms involving the compensators alone are
\begin{align}
S_{WG} &= \int \rd^4x\, \rd^2\theta\, \rd^2\btheta\, E\, \Bigg\{
     \frac{1}{4} L^2 - \frac{1}{2} \eta\bar\eta - 2 \xi L V - \chi \bar\chi
     \eol & \quad
     + \frac{1}{4} \bar\gamma \eta^2 + \frac{1}{4} \gamma \bar \eta^2
     + \bar\gamma \bar w \eta \chi + \gamma w \bar\eta \bar\chi
     - \frac{1}{R} W^\alpha W_\alpha
     \Bigg\}~.
\end{align}
The terms involving mixing between the vector and 
the supergravity multiplets are given by
\begin{align}
S_{WH} &= \int \rd^4x\, \rd^2\theta\, \rd^2\btheta\, E\, \Bigg\{
     - \hat U (w \bar \chi + \bar w \chi)
     + 2\ri \bar w \Psi^\alpha W_\alpha
     - 2\ri w \bar \Psi_\dalpha\bar W^\dalpha
     \eol & \quad
     - \frac{\ri}{3} H^{\dalpha \alpha} \CD_{\alpha \dalpha} (\bar w \chi - w \bar \chi) \Bigg\}~,
\end{align}
while those involving the tensor and supergravity multiplets are
\begin{align}
S_{G H} &= \int \rd^4x\, \rd^2\theta\, \rd^2\btheta\, E\, \Bigg\{
     \frac{1}{2} \hat U (\gamma \bar\eta + \bar\gamma \eta)
     - \frac{\ri}{2} \gamma L \CD^\alpha \Psi_\alpha
     + \frac{\ri}{2} \bar\gamma L \bar\CD_\dalpha \bar \Psi^\dalpha
     \eol & \quad
     - \frac{\ri}{3} H^{\dalpha \alpha} \CD_{\alpha \dalpha}
     \left(\bar\gamma \eta - \gamma \bar\eta\right) \Bigg\}~.
\end{align}
The pure supergravity sector yields
\begin{align}
S_{HH} &= \int \rd^4x\, \rd^2\theta\, \rd^2\btheta\, E\, \Bigg\{
     - \frac{3}{4} \hat U^2
     - \frac{1}{16} H^{\dalpha \alpha} \nabla^\beta (\bar\nabla^2 - 4R) \nabla_\beta H_{\alpha \dalpha}
     \eol & \quad
     + \frac{1}{48} ([\nabla_\alpha, \bar\nabla_\dalpha] H^{\dalpha \alpha})^2
     - \frac{1}{4} (\nabla_{\alpha \dalpha} H^{\dalpha \alpha})^2
     - \frac{1}{4} R \bar R H^{\dalpha \alpha} H_{\alpha \dalpha}
     \eol & \quad
     - \Psi^\alpha \bar\nabla_\dalpha \CD_\alpha \bar \Psi^\dalpha
     - \frac{1}{4} (\gamma \nabla^\alpha \Psi_\alpha - \bar\gamma \bar\nabla_\dalpha \bar \Psi^\dalpha)^2
     -\frac{\bar w^2}{4} \Psi^\alpha \bar\nabla^2 \Psi_\alpha
     -\frac{w^2}{4} \bar \Psi_\dalpha \CD^2 \bar \Psi^\dalpha \Bigg\}.
\end{align}

As in the Minkowski case, the superfield $\hat U$ is an auxiliary and may
be integrated out algebraically. Doing so, we find that only a certain combination
of chiral superfields survives, which can be denoted
\begin{align}
\bar\varphi \phi := \frac{1}{3} \bar w \chi + \frac{1}{3} \bar\gamma \eta
\end{align}
We interpret $\varphi$ as a background phase, corresponding to the background
value of a chiral compensator $\Phi$, while $\phi$ is its quantum deformation.
In our previous paper \cite{n2_sugra}, we denoted the combination
$\bar\varphi \phi$ by $\sigma$; to make the analogy with $\N=2$ as strong
as possible, we have restored a background value to this compensator.

We arrive at two decoupled actions $S = S_{\rm old} + S_\Psi$.
The first is the linearized old minimal supergravity action in AdS
\begin{align} \label{4.70}
S_{\rm old} &=  -\int \rd^4x\, \rd^2\theta\, \rd^2\btheta\, E\, \Bigg\{
      \frac{1}{16} H^{\dalpha \alpha} \nabla^\beta (\bar\nabla^2 - 4R) \nabla_\beta H_{\alpha \dalpha}
     - \frac{1}{48} ([\nabla_\alpha, \bar\nabla_\dalpha] H^{\dalpha \alpha})^2
     \eol & \quad
     + \frac{1}{4} (\nabla_{\alpha \dalpha} H^{\dalpha \alpha})^2
     + \frac{R \bar R}{4} H^{\dalpha \alpha} H_{\alpha \dalpha}
     + \ri H^{\dalpha \alpha} \nabla_{\alpha \dalpha} (\bar\varphi \phi - \varphi \bar\phi)
     + 3 (\phi \bar\phi - \bar\varphi^2 \phi^2 - \varphi^2 \bar\phi^2)\Bigg\} 
\end{align}
which is invariant under the gauge transformations
\begin{subequations}\label{eq_N1gaugeOldMin2}
\begin{align}
\delta H_{\alpha \dalpha} &= \nabla_\alpha \bar L_\dalpha - \bar \nabla_\dalpha L_\alpha ~,
\label{eq_N1gaugeOldMin2-H}
\\
\delta \phi &= - \frac{\varphi}{12} (\bar \nabla^2 - 4 R) \nabla^\alpha L_{\alpha}~.
\end{align}
\end{subequations}
The other sector is the massless gravitino action in AdS
\begin{align}
S_\Psi &=   \int \rd^4x\, \rd^2\theta\, \rd^2\btheta\, \, E\, \Bigg\{
    -  \Psi^\alpha \bar\nabla_\dalpha \nabla_\alpha \bar \Psi^\dalpha
     -\frac{\bar w^2}{4} \Psi^\alpha \bar\nabla^2 \Psi_\alpha
     -\frac{w^2}{4} \bar \Psi_\dalpha \nabla^2 \bar \Psi^\dalpha
     \eol & \quad
     - \frac{1}{4} (\gamma \nabla^\alpha \Psi_\alpha - \bar\gamma \bar\nabla_\dalpha \bar \Psi^\dalpha)^2
     + 2\ri \bar w \Psi^\alpha W_\alpha
     - 2\ri w \bar \Psi_\dalpha\bar W^\dalpha
     \eol & \quad
     - \frac{\ri}{2} L ( \gamma  \nabla^\alpha \Psi_\alpha
     - \bar\gamma  \nabla_\dalpha \bar \Psi^\dalpha )
     + \frac{1}{4} L^2 - 2 \xi L V 
     - \frac{1}{R} W^\alpha W_\alpha \Bigg\}~,
\end{align}
where $W_\alpha \equiv \frac{1}{4} (\bar \nabla^2 - 4 R) \nabla_\alpha V$.
Its gauge symmetries are described by
\begin{subequations}
\begin{align}
\delta \Psi_\alpha &= \nabla_\alpha \Omega + \Lambda_\alpha ~,\\
\delta V &= -\ri \bar w\Omega + \ri w \bar \Omega + \l + \bar\l ~,\\
\delta L &= \ri\gamma \nabla^\alpha \Lambda_\alpha - \ri \bar\gamma \bar\nabla_\dalpha \bar\Lambda^\dalpha~,
\end{align}
\end{subequations}
where $\l$ is the covariantly chiral gauge parameter associated with
the usual gauge invariance of an abelian vector multiplet.

The supergravity action (\ref{4.70}) reduces to (\ref{eq_N1OldMin}) for $\vf =1$.
In the rigid supersymmetric limit, the gravitino multiplet action $S_\Psi$
correctly reduces to its flat superspace counterpart  derived in \cite{n2_sugra}
(see also \cite{LR}).

\subsection{Dual versions of the gravitino multiplet action}
The gravitino multiplet action $S_\Psi$ which we have found is quite interesting since
the coupling between the tensor and vector multiplets allows several duality
transformations (compare with \cite{Kuzenko:2004tn}). For example, 
dualizing the linear multiplet to a chiral multiplet gives
\begin{align}
\tilde S_\Psi &= \int \rd^4x\, \rd^2\theta\, \rd^2\btheta\, \, E\, \Bigg\{
     - \Psi^\alpha \bar\nabla_\dalpha \nabla_\alpha \bar \Psi^\dalpha
     -\frac{\bar w^2}{4} \Psi^\alpha \bar\nabla^2 \Psi_\alpha
     -\frac{w^2}{4} \bar \Psi_\dalpha \nabla^2 \bar \Psi^\dalpha
     \eol & \quad
     + 2\ri \bar w \Psi^\alpha W_\alpha
     - 2\ri w \bar \Psi_\dalpha\bar W^\dalpha
     - 2 \ri \x ( V + \phi + \bar\phi)
               (\gamma \nabla^\alpha \Psi_\alpha - \bar\gamma \bar\nabla_\dalpha \bar \Psi^\dalpha)
     \eol & \quad
     - 4 \x  ( V + \phi + \bar \phi)^2
     - \frac{1}{R} W^\alpha W_\alpha \Bigg\}~,
\end{align}
where $W_\alpha \equiv \frac{1}{4} (\bar \nabla^2 - 4 R) \nabla_\alpha V$.
The gauge invariances are
\begin{subequations}
\begin{align}
\delta \Psi_\alpha &= \nabla_\alpha \Omega + \Lambda_\alpha ~,\\
\delta V &= -\ri \bar w\Omega + \ri w \bar \Omega + \l + \bar\l ~,\\
\delta \phi &= +\frac{\ri \bar\gamma}{4} (\bar\nabla^2 - 4 R) \bar \Omega
     - \l~.
\end{align}
\end{subequations}
We have written the action to make it obvious that $\phi$ is a Stueckelberg field;
it may be eliminated by sacrificing the gauge invariance $\l$. Similarly, one may
eliminate $V$ by employing the $\Omega$ gauge invariance.

We may also perform a duality on the vector multiplet to give
\begin{align}
\tilde S'_\Psi &= \int d^{8}z\, E\, \Bigg\{
     - \Psi^\alpha \bar\nabla_\dalpha \nabla_\alpha \bar \Psi^\dalpha
     + \frac{\bar w^2}{4} \frac{\bar \alpha}{\alpha} \Psi^\alpha \bar\nabla^2 \Psi_\alpha
     + \frac{w^2}{4} \frac{\alpha}{\bar\alpha} \bar \Psi_\dalpha \nabla^2 \bar \Psi^\dalpha
     \eol & \quad
     + \frac{1}{4} \left(L - \ri \gamma \nabla^\alpha \Psi_\alpha
     + \ri \bar\gamma \bar\nabla_\dalpha \bar\bar \Psi^\dalpha \right)^2
     \eol & \quad
     + \frac{2\xi^2}{\alpha R} \left(\chi^\alpha + \frac{\ri}{\xi} W^\alpha - \ri \gamma \Psi^\alpha \right)
          \left(\chi_\alpha + \frac{\ri}{\xi} W_\alpha - \ri \gamma \Psi_\alpha \right)
     \eol & \quad
     + \frac{2\xi^2}{\bar\alpha \bar R} \left(\bar\chi_\dalpha - \frac{\ri}{\xi} \bar W_\dalpha
          + \ri \bar\gamma \bar \Psi_\dalpha \right)
          \left(\bar\chi^\dalpha - \frac{\ri}{\xi} \bar W^\dalpha + \ri \bar\gamma \bar \Psi^\dalpha \right)
     \Bigg\}~.
\end{align}
The parameter $\alpha$ is complex and constrained to have unit
real part. The resemblance to the dual $\N=2$ action \eqref{eq_N2Dualaction}
is strong; likely one can derive the former from the latter.
This action is invariant under the gauge transformations
\begin{subequations}
\begin{align}
\delta \Psi_\alpha &= \nabla_\alpha \Omega + \Lambda_\alpha~, \\
\delta \chi_\alpha &= \ri \gamma \Lambda_\alpha + \ri \phi_\alpha~, \\
\delta W_\alpha &= -\xi \phi_\alpha
     - \frac{1}{8} (\bar\nabla^2 - 4 R)\nabla_\alpha (\bar\alpha \bar w \Omega + \alpha w \bar\Omega)~,
\end{align}
\end{subequations}
where $\phi_\alpha$ and $W_\alpha$ are both reduced chiral spinors. We have written
the action in a way which makes clear that $W_\alpha$ is a Stueckelberg field and can
be removed by fixing the $\phi_\alpha$ gauge degree of freedom. Similarly, one may eliminate
$\chi_\alpha$ by employing the $\Lambda_\alpha$ gauge invariance.

\section{Supercurrent}
We are now in a position to postulate the general supercurrent conservation equation
for arbitrary matter systems coupled to $\N=2$ supergravity with vector and tensor compensators. 
We have confirmed the structure of the linearized supergravity gauge transformations in
an AdS background, eq. (\ref{eq_AdsGauge}).
Moreover, it has been argued in subsection 3.1 that the same transformation laws 
hold in an \emph{arbitrary} supergravity background. 
We can therefore state the gauge transformation laws of the supergravity prepotentials: 
\begin{subequations}\label{eq_N2Gauge}
\begin{align}
{\delta} \J &= 4 \bar\Delta(\bar\qOmega^{ij} \cG_{ij}) ~,\\
{\delta} \cV^{ij} &= -4\qOmega^{ij} \bar \cW - 4 \bar \qOmega^{ij} \cW ~, \\
{\d} \cH &= (\cD^{ij} + 4 S^{ij}) \qOmega_{ij} + 
     (\bar\cD^{ij} + 4 \bar S^{ij}) \bar \qOmega_{ij}~.
\end{align}
\end{subequations}
This information is sufficient  to identify the relevant  supercurrent multiplet.

\subsection{Main construction} 
Given a matter system coupled to $\cN=2$ supergravity, we  give small disturbances,
$\bf H$, ${\bf V}^{ij}$ and $\bf \J$, 
to the gravitational superfield $\cH$ and the compensators $\cV^{ij}$ and $\J$, 
respectively, the latter being defined 
as in eqs. (\ref{eq_WMezin}) and (\ref{eq_Gprepotential}).
To first order, the matter action changes as
\begin{align}
S^{(1)} &= \int \rd^4x\, \rd^4\theta\, \rd^4\bar\theta \, E\, \Big\{
     \cJ \qH + \cT_{ij} \qV^{ij}
     \Big\}
     + \left\{  \int \rd^4x\, \rd^4\theta\, \cE\, \cY \qPsi +{\rm c.c.} \right\}
     ~,
\label{5.2}
\end{align}
where 
\bea
\cJ = \frac{{ \d} S}{{ \d}\cH }~, \qquad \cT_{ij} = \frac{{\d} S}{{\d} \cV^{ij} }~, 
\qquad \cY = \frac{{\d} S}{{\d} \J }~.
\eea
The sources $\cJ$ and $\cT_{ij}$ must be real,  and $\cY$ covariantly  chiral.
In addition, $\cY$ and $\cT_{ij}$ must obey the constraints
\begin{subequations}
\begin{gather}\label{eq_CurrentConstraints}
\cD_\alpha{}^{(k} \cT^{ij)} = \bar\cD_\dalpha{}^{(k} \cT^{ij)} = 0 ~,\\
(\cD^{ij} + 4 S^{ij})\cY = (\bar \cD^{ij} + 4 \bar S^{ij})\bar\cY~,
\end{gather}
\end{subequations}
due to the gauge invariances of $\qV^{ij}$ and $\qPsi$ given by eqs. 
(\ref{pre-gauge1}) and (\ref{eq_PsiGauge}).
In the case that the disturbances $\bf H$, ${\bf V}^{ij}$ and $\bf \J$
correspond to a supergravity gauge transformation, eq. (\ref{eq_N2Gauge}), 
the functional  $S^{(1)}$ must vanish provided the matter fields are placed on the mass shell.
Since the gauge parameters ${\bf \O}^{ij}$ are unconstrained,  we end up with 
 the conservation equation
\begin{align}\label{eq_N2current}
\frac{1}{4} (\BCD_{ij} + 4 \bar S_{ij}) \cJ = \cW \cT_{ij} - \cG_{ij} \cY~.
\end{align}

There is an alternative way to derive the conservation equation (\ref{eq_N2current}).
We can start from $\cN=2$ supergravity without matter and choose an on-shell background,  
that is a solution to  the supergravity equations of motion (\ref{2.11}).
We then linearize the supergavity action around the background chosen. 
The linearized supergravity action,  $S^{(2)}$, must be invariant under the 
supergravity gauge transformations (\ref{eq_N2Gauge}), 
as well as under the gauge invariances of $\qV^{ij}$ and $\qPsi$ given by eqs. 
(\ref{pre-gauge1}) and (\ref{eq_PsiGauge}).
The linearized supergravity multiplet can be further coupled to external sources
by replacing $S^{(2)} \to S^{(2)} +S^{(1)}$, with $S^{(1)}$ given by eq. (\ref{5.2}).
The obtained action is gauge invariant provided the 
conservation equation (\ref{eq_N2current}) holds.
 
If the supergravity compensators obey the equation 
$\cW \bar \cW = \cG$, which is one of the equations of motion for pure supergravity
(\ref{2.11}), then $\cW$ and $\cG_{ij}$ may be fixed to to be constant
by applying appropriate super-Weyl and local ${\rm U(2)_R}$ transformations.
However, the presence of the matter will modify the
supergravity equation of motion (even though the matter is on-shell),
so we cannot assume $\cW \bar \cW = \cG$.
This eliminates the possibility of simultaneously gauging both compensators
to be constant (or even covariantly constant). However, it is often the
case that a theory couples to only one of the two compensators, and then
it is quite natural to choose the gauge where $\cG$ or $\cW \bar \cW$ is
constant, as appropriate.

Before elaborating on the supercurrent (\ref{eq_N2current}), we should first review
the $\N=1$ situation, paying particular attention to the case in AdS
where additional restrictions on the supercurrent arise. Then we consider
some consequences of this proposed $\N=2$ supercurrent in a Minkowski
background. Finally, we discuss it in a more general supergravity background
and present some examples of its application.

\subsection{$\N=1$ supercurrents}

Textbook derivations \cite{GGRS,BK}
of the supercurrent multiplet in the old minimal formulation for $\cN=1$ 
supergravity make use of the chiral prepotential $\F$ and its conjugate $\bar \F$
introduced originally by Siegel and Gates
\cite{S,SG}. These objects and the  gravitational superfield $\cH^m$ \cite{OS,S} 
are the only prepotentials, modulo purely gauge degrees of freedom,  
in terms of which the Wess-Zumino constraints \cite{WZ} 
are solved \cite{S}. In this paper, we work with a super-Weyl invariant extension 
of old minimal supergravity, which makes use of a covariantly chiral scalar superfield
$\vf$  and its conjugate in addition to the Weyl multiplet described by the covariant derivatives $\nabla_A$. 
Because of the super-Weyl symmetry, which leaves the chiral combination $( \F \vf )$ invariant,
there is no need at all to resort to $\F$ .
However, there appear some new technical nuances regarding the derivation of the supercurrent 
conservation equation. 
Such a derivation is similar to the $\cN=2$ construction described above.

Given a matter system coupled to old minimal supergravity, we  give small disturbances,
$H^m$ and $\f$, 
to the gravitational superfield $\cH^m$ and the compensator $\vf$, 
respectively. To first order, the matter action changes by 
\begin{align}
S^{(1)} = \int \rd^4x\, \rd^2\theta\, \rd^2\bar\theta \, E\, J^{\dalpha \alpha} H_{\alpha \dalpha}
     - \left\{ 3 \int \rd^4x\, \rd^2\theta\, \cE\, X \phi +{\rm c.c.}\right\}
\label{5.6}
\end{align}
where\footnote{The factor of 3 in (\ref{5.6}) is introduced for convenience.} 
\bea
J_{\a\ad}  = \frac{ { \d} S}{ { \d} \cH^{\a\ad }}~, \qquad
X = -\frac{1}{3} \frac{\d S}{\d \vf }~.
\eea
By construction, $X$ is covariantly chiral. 
If $H_{\a\ad}$ and $\f$ correspond to a supergravity gauge transformation, 
then $S^{(1)}$ must vanish with the matter fields on-shell. 
In the case that $\vf$ is chosen to be annihilated by the spinor covariant derivatives, 
the $\N=1$ supergravity transformations are described by eq. \eqref{eq_N1gaugeOldMin2}. 
In the general case, the transformation of $\f$  is as follows\footnote{This can be
justified in a number of ways. In analogy to our $\N=2$ argument, this is
the unique possibility (up to overall normalization) which transforms covariantly
under super-Weyl transformations when $\varphi$ has weight 1 and $L_\alpha$ has weight
-3/2. It may also be constructed by accompanying any coordinate transformation
with an appropriate super-Weyl transformation to fix the chiral prepotential $\s$
hidden within the covariant derivatives. Alternatively, it may also be derived
explicitly in conformal superspace \cite{Butter:2009wy} and then gauge fixed to
Poincar\'e superspace.}
\begin{align}
\delta \phi = -\frac{1}{4} (\bar\nabla^2 - 4 R) (L^\alpha \nabla_\alpha \varphi)
     -\frac{\varphi}{12} (\bar\nabla^2 - 4 R) \nabla^\alpha L_\alpha~.
     \label{5.8}
\end{align}
Of course, due to the super-Weyl invariance, we can always choose a super-Weyl gauge 
$\varphi \bar\varphi ={\rm const}$, and then the transformation (\ref{5.8}) reduces to 
 \eqref{eq_N1gaugeOldMin2}. 
The condition $S^{(1)}=0$ for $H_{\a\ad}$ and $\f$ given by eqs. (\ref{eq_N1gaugeOldMin2-H})
and (\ref{5.8}), respectively, is the conservation equation \cite{Butter:2010hk}
\begin{align}
\bar \nabla^\dalpha J_{\alpha \dalpha} = \varphi \nabla_\alpha X - 2X \nabla_\a\vf 
~.
\end{align}
If $\nabla_\a \vf =0$, then this equation reduces, modulo a trivial rescaling,  to eq. (\ref{AdS-mcl}).

The above argument is easily generalized in the presence of other compensating multiplets.
Essentially the only input needed is the generalization of the rule (\ref{5.8})
for the supergravity gauge transformation of those multiplets.
For example, if there is a single linear multiplet $\cL$ acting as a compensator,
as in new minimal supergravity, one may construct its `quantum' variation $L$
out of a chiral spinor $\phi_\alpha$,
\begin{align}
\cL \rightarrow \cL + L~, \qquad L = \nabla^\alpha \phi_\alpha + \bar\nabla_\dalpha \bar\phi^\dalpha ~,
\end{align}
with $\phi_\alpha$ transforming as
\begin{align}
\delta \phi_\alpha = \frac{1}{4} (\bar\nabla^2 - 4 R) (L_\alpha \cL)
\end{align}
under the supergravity gauge transformation. 
If $\phi_\alpha$ occurs only in the combination
$L$, then it possesses the additional symmetry
$\delta \phi_\alpha = \ri W_\alpha$, where $W_\alpha$ is a reduced chiral spinor.

Any matter action coupled to one or both of these compensators
must be invariant under the supergravity gauge transformations.
Defining the new trace multiplet $\chi_\alpha$ by its contribution
to $S^{(1)}$,
\begin{align}
\int \rd^4x\, \rd^2\theta\, \cE\, \chi^\alpha \phi_\alpha~,\qquad
\nabla^\alpha \chi_\alpha - \bar\nabla_\dalpha \bar\chi^\dalpha = \bar \nabla_\dalpha \chi_\alpha = 0~,
\end{align}
leads to the modified conservation equation
\begin{align}\label{eq_AdS_KS_1}
\bar \nabla^\dalpha J_{\alpha \dalpha} = \cL \chi_\alpha + \varphi \nabla_\alpha X
-2 X \nabla_\a \vf
~.
\end{align}

Suppose we have chosen the super-Weyl gauge
$\varphi \bar\varphi = \textrm{const.}$ Then $\varphi$ is covariantly
constant and can be made truly constant by fixing the $\rm U(1)_R$ gauge.
The question remains whether it is possible for $\cL$ to be constant as well.
However, it is quite clear that this cannot be the case in general since the
background linear compensator obeys the condition
\begin{align}
(\bar\nabla^2 - 4 R) \cL = (\nabla^2 - 4 \bar R) \cL = 0~,
\end{align}
and so one cannot simultaneously choose to work in an AdS frame $R = \mu \neq 0$
\emph{and} choose the background compensator $\cL$ to be constant.

To put it another way, if we naively generalize the flat space supercurrent (\ref{S-multiplet}),
ignoring the background compensators (i.e. assuming that they are
unity), we find
\begin{align}\label{eq_AdS_KS_2}
\bar \nabla^\dalpha J_{\alpha \dalpha} = \chi_\alpha + \nabla_\alpha X~.
\end{align}
Applying $(\bar \nabla^2 - 4 R)$ to both sides, we find immediately that
\begin{align}
0 = (\bar \nabla^2 - 4 R) \chi_\alpha = -4 R \chi_\alpha~,
\end{align}
which is consistent only if $R$ vanishes.

So the naive supergravity extension \eqref{eq_AdS_KS_2}
of the Komargodski-Seiberg supercurrent, eq. (\ref{S-multiplet}), 
is inconsistent. The consistent extension is given by the
relation (\ref{eq_AdS_KS_1}).
The key point is the compensator $\cL$ must be taken seriously as a
superfield with nontrivial coordinate dependence.

\subsection{$\N=2$ supercurrents in Minkowski background}
We consider next some aspects of the $\N=2$ supercurrent \eqref{eq_N2current}
in a Minkowski background, meaning that we take all of the $\N=2$
torsion superfields to vanish while simultaneously taking the background
compensators to be constant, $ \cW = w ={\rm const}$ and 
$\cG^{ij} = g^{ij} ={\rm const}$.
In this case, the $\N=2$ supercurrent conservation equation reads
\begin{align}
\frac{1}{4} \bar D_{ij} \cJ = w \cT_{ij} - g_{ij} \cY~.
\end{align}
(In our previous work \cite{n2_sugra}, we fixed $w=\ri$.)

One particularly interesting application of this equation, which we
neglected to discuss in \cite{n2_sugra}, is that it naturally leads to the 
$\N=1$ supercurrent (\ref{S-multiplet})
discussed by Komargodski and Seiberg \cite{KS2}.
Defining the $\N=1$ supercurrent $J_{\alpha \dalpha}$ by the rule \cite{KT}
\begin{align}
J_{\alpha \dalpha} :=
     \frac{1}{4} [D_\alpha{}^\2~, \bar D_{\dalpha \2}] \cJ\vert
     - \frac{1}{12} [D_\alpha{}^\1~, \bar D_{\dalpha \1}] \cJ\vert~,
\end{align}
one can show it obeys
\begin{align}
\bar D^\dalpha J_{\alpha \dalpha} = \chi_\alpha + D_\alpha X
\end{align}
where the contributions to the trace multiplet are given by
\begin{align}
\chi_\alpha &:= 4\ri g_{\1\2} \cY_\alpha, \qquad
X := \frac{1}{3} w T_{\1\1} + \frac{2}{3} g_{\2\2} Y \\
T_{\1\1} &:= \cT_{\1\1} \vert~, \qquad
Y := \cY \vert~, \qquad
\cY_\alpha := \frac{\ri}{2} D_\alpha{}^\2 \cY \vert~.
\end{align}
(One should keep in mind that $\ri g_{\1\2}$ is real.)
Because of the constraints \eqref{eq_CurrentConstraints},
$T_{\1\1}$ and $Y$ are both chiral superfields while $\cY_\alpha$ is a chiral
field strength obeying the Bianchi identity
$D^\alpha \cY_\alpha = \bar D_\dalpha \bar \cY^\dalpha$.

Particularly illuminating is the case where $\cT_{ij}$ vanishes
(i.e. the $\N=2$ theory couples only to the tensor compensator).
Then we have
\begin{align}
\chi_\alpha &= 4\ri g_{\1\2} \cY_\alpha, \qquad
X = \frac{2}{3} g_{\2\2} Y.
\end{align}
The background values $g_{\2\2}$ and $g_{\1\2}$ rotate under
the global $\rm SU(2)_R$ symmetry; for a certain choice of gauge,
either one (but not both) can be eliminated. In these situations,
we clearly see the supercurrent associated with old minimal 
($g_{\1\2} = \chi_\alpha = 0$) or new minimal supergravity
($g_{\2\2} = X = 0$); in the general case, the current has
the form  (\ref{S-multiplet}).
A global $\rm SU(2)_R$ rotation connects them all.

\subsection{Supercurrent in a supergravity background: Examples}
We conclude this section by giving several examples of supercurrents
in curved superspace.

\subsubsection{Abelian gauge theory with a Fayet-Iliopoulos term}
Consider an abelian $\cN=2$ gauge theory coupled to the tensor compensator $\cG^{ij}$:
\begin{align}
S = \frac{1}{2} \int \rd^4x\, \rd^4\theta\, \cE\, W^2 
     -\l  \int \rd^4x\, \rd^4\theta\, \rd^4\bar\theta\, E\, \cG^{ij} V_{ij}~.
\end{align}
As usual,  $W$ denotes the covariantly chiral field strength of the $\cN=2$ vector multiplet, 
and $V_{ij}$ the corresponding Mezincescu prepotential.
If we impose the super-Weyl and $\rm SU(2)_R$ gauges where $\cG^{ij}$ is constant,
the tensor compensator coupling resembles a Fayet-Iliopoulos term.
In addition to the usual supercurrent
\begin{align}
\cJ = W \bar W~,
\end{align}
we have nontrivial coupling due to the tensor compensator
\begin{align}
\cY = - \lambda W~,
\end{align}
while $\cT_{ij} = 0$ since the model does not couple to the vector compensator.
It is easy to see that the supercurrent equation \eqref{eq_N2current} holds due
to the equation of motion
\begin{align}
\frac{1}{4} (\cD^{ij} + 4 S^{ij}) W = \frac{1}{4} (\bar\cD^{ij} + 4 \bar S^{ij}) \bar W
     = \l \cG^{ij}~.
\end{align}

\subsubsection{Supersymmetric Yang-Mills theory} 
We next consider a general $\cN=2$ supersymmetric  Yang-Mills (SYM) theory with a hypermultiplet
transforming in some representation of the gauge group. The simplest off-shell description of this theory 
in the presence of supergravity 
is provided by the projective-superspace formulation for $\cN=2$ supergravity-matter systems
developed in \cite{KLRT-M1,KLRT-M2}. The action is 
\begin{align}
S_{\rm YM} =\frac{1}{2g_{\rm YM}^2} \int \rd^4x\, \rd^4\theta\, \cE\, \mathrm{Tr}\, ({\bm \cW}^2)
+  \frac{\ri}{2\pi}\oint_C  v^i \rd v_i
     \int \rd^4 x \,{\rm d}^4\theta {\rm d}^4{\btheta}
     \,E\, \frac{\cW \bar\cW}{ ( \Sigma^{++} )^2}\, \breve{\U}^+ \U^+~,~~~~~
\label{5.27}
\end{align}
with $g_{\rm YM}$ the coupling constant.
Here the first term describes the pure SYM sector, with $\bm \cW$ the gauge-covariantly 
chiral field strength, 
\bea
\bar{\bm \cD} ^\ad_i {\bm \cW}= 0~, \quad
4{\bm \S}^{ij} :=  \Big({\bm \cD}^{ij}+4S^{ij}\Big) {\bm \cW }&=& 
\Big( \bar {\bm \cD}^{ij}  +4\bar{S}^{ij}\Big)\bar{\bm \cW} ~.
\eea
The gauge-covariant derivatives ${\bm \cD}_A = ({\bm \cD}_a, {\bm \cD}_\a^i , \bar {\bm \cD}_i^\ad )$
differ from the supergravity covariant derivatives $\cD_A$, which are described in Appendix A, 
by the presence of (i) the Yang-Mills connection, and (ii) the U(1) connection associated with 
the vector multiplet compensator $\cW$. The spinor derivatives obey the anti-commutation relations
\begin{subequations}
\begin{align}
\{{\bm\cD}_\alpha^i, {\bm\cD}_\beta^j\} &= \{{\cD}_\alpha^i, {\cD}_\beta^j\}
     -2\ri \,\eps^{ij} \eps_{\alpha \beta}
     \left(\bar \cW \hat e + \bar{\bm{\cW}}\right)~, \\
\{\bar{\bm\cD}^\dalpha_i, \bar{\bm\cD}^\dbeta_j\} &=
     \{\bar{\cD}^\dalpha_i, \bar{\cD}^\dbeta_j\} 
     +2\ri \,\eps_{ij} \eps^{\dalpha \dbeta}
     \left(\cW \hat e + {\bm{\cW}}\right)~,
\end{align}
\end{subequations}
with $\hat e$ the $\rm U(1)$ charge operator associated with $\cW$.
The anticommutator $\{{\bm\cD}_\alpha^i, \bar {\bm\cD}^\dbeta_j\}$ has the same
form \eqref{a-c2} as $\{{\cD}_\alpha^i, \bar{\cD}^\dbeta_j\}$
with the replacement of $\cD_c$ with ${\bm\cD}_c$.

The second term in (\ref{5.27}) is the hypermultiplet action.
It involves a closed contour in an auxiliary  
isotwistor  variable $v^i \in  {\mathbb C}^2 \setminus  \{0\}$, with respect to which 
the {\it arctic} hypermultiplet $\U^+ (z,v)$, its {\it smile}-conjugate $\breve{\U}^+(z,v)$
and $\S^{++}(z,v):=\S^{ij}(z) v_iv_j$ are holomorphic homogeneous functions, 
with $\S^{ij}$ defined in (\ref{1.1}). The hypermultiplet $\U^+ $ and  its {\it smile}-conjugate $\breve{\U}^+$
are special covariant projective  supermultiplets\footnote{See \cite{KLRT-M2} for the general properties
of the covariant projective supermultiplets  and their supergravity gauge transformation laws.} 
 annihilated by half of the supercharges. 
Specifically, $\U^+$ obeys the constraints 
\be
{\bm \cD}^+_{\a} \U^{+}  = {\bar {\bm \cD}}^+_{\ad} \U^{+}  =0~, \qquad \quad 
{\bm \cD}^+_{ \a}:=v_i\, {\bm \cD}^i_{ \a} ~, \qquad
{\bar {\bm \cD}}^+_{\dot  \a}:=v_i\,{\bar {\bm \cD}}^i_{\dot \a} ~,
\label{ana-introduction}
\ee  
and similar constraints hold for $\breve{\U}^+$.
In terms of the complex variable $\z$ defined using $v^i =(v^{\1}, v^{\2}) = v^{1} (1,\z)$,
the explicit functional forms of $\U^+(v) $ and $\breve{\U}^+$ are
\bea
\U^+(v) = v^{\1} \sum_{n=0}^{\infty} \U_n \z^n~, \qquad
\breve{\U}^+ (v) &=&  v^{\2}  \sum_{n=0}^{\infty} (-1)^n {\bar \U}_n 
\frac{1}{\z^n}~.
\eea
The hypermultiplet action in (\ref{5.27}) is invariant under arbitrary re-scalings of $v^i$, 
and therefore these variables are homogeneous coordinates for ${\mathbb C}P^1$.
The arctic hypermultiplet is assumed to be charged under the U(1) gauge group associated with 
the vector multiplet compensator, and we denote by  $e$ the charge of $\U^+$.

\newcommand{\ymW}{{\boldsymbol {\mathcal W}}}

It can be shown that the hypermultiplet equations of motion are equivalent to 
\bea
\U^+(v) = \U^i v_i~, 
\eea
where $\U^i$ is an ordinary isospinor superfield obeying  
the covariant constraints
\begin{align}
{\bm \cD}_\alpha^{(i} \U^{j)} = \bar {\bm \cD}_\dalpha^{(i} \U^{j)} = 0~.
\label{5.32}
\end{align}
These constraints prove to imply that the multiplet $\U^i$ is on-shell, 
and its mass $m$ is equal to $ |e|$.
In the rigid supersymmetric case, the constraints were introduced by Sohnius \cite{Sohnius:1978fw}.
It can  also be shown that the equation of motion for the SYM multiplet implies that 
\bea
\frac{1}{g_{\rm YM}^2}  \mathrm{Tr\,}(\ymW {\bf \Sigma}^{ij}) = \ri \bar \U^{(i} \ymW \U^{j)} ~.
\eea
Now, if we define the supercurrent as in \cite{KT}, 
\begin{align}
\cJ =  \frac{1}{g_{\rm YM}^2} \mathrm{Tr\,} (\ymW \bar \ymW) -\frac{1}{2} \bar \U_k \U^k~,
\end{align}
a simple calculation shows that
 $\cJ$ indeed satisfies the conservation equation  \eqref{eq_N2current} with $\cY =0$ and 
 \begin{align}
\cT_{ij} = - \ri e \bar \U_{(i} \U_{j)}~.
\end{align}
We have $\cY=0$ since the SYM action is independent of the tensor compensator.

\section{Conclusion}
For the minimal off-shell $\cN=2$ supergravity  with vector and tensor compensators \cite{deWPV},
we have  constructed the linearized action in the AdS background, 
eq.  \eqref{eq_N2action}.  A main advantage of our construction is that it
has revealed the gauge transformations of the supergravity prepotentials, 
eq.   \eqref{eq_N2Gauge}, and uncovered the supercurrent multiplet  \eqref{eq_N2current}
corresponding to those matter theories  which couple to the supergravity chosen. 
The $\cN=2$ supergravity  formulation with vector and tensor compensators
allows one to realize a huge class of matter couplings within the projective-superspace
approach developed in  \cite{KLRT-M1, KLRT-M2, K-dual08}.

The action \eqref{eq_N2action}, although constructed in AdS, could serve as a springboard
to reconstructing the linearized supergravity action in an arbitrary on-shell background. The only
contributions missing are those arising from $W_{\alpha \beta}$, the $\N=2$
analogue of the $\N=1$ superfield $W_{\alpha \beta \gamma}$, which contains
the conformal Weyl tensor and which vanishes in an AdS background but not in
a general on-shell background. These are, in principle, straightforward to restore
by including terms involving $W_{\alpha \beta}$ in the linearized action
and determining their coefficients by requiring supergravity gauge invariance.
The result is just a first step toward quantizing pure $\N=2$ supergravity
and then performing one-loop calculations (similar to what was done in
$\cN=1$ supergravity by Grisaru and Siegel \cite{Grisaru:1981xm}).
Since $\cN=2$ superfield supergravity is a reducible gauge theory, its covariant quantization
is nontrivial.

Another interesting open problem would be to construct massive extensions
of the linearized $\N=2$ supergravity action in AdS. 
Even in the super-Poincar\'e case, such a problem has not been addressed.

\noindent
{\bf Acknowledgements:}\\
This work is supported in part by the Australian Research Council 
and by a UWA Research Development Award.

\appendix

\section{$\cN=2$ superconformal geometry}
\label{Appendix A}
We give a summary of the superspace geometry for $\N=2$
conformal supergravity which was originally  introduced in \cite{Howe},
as a generalization of \cite{Grimm}, and later elaborated in \cite{KLRT-M2}.
A curved four-dimensional  $\cN=2$ superspace  $\cM^{4|8}$ is parametrized by
local  coordinates  $z^{{M}}=(x^{m},\q^{\mu}_\imath,{\bar \q}_{\dot{\mu}}^\imath)$,
where $m=0,1,\cdots,3$, $\mu=1,2$, $\dot{\mu}=1,2$ and  $\imath= \1,\2$.
The Grassmann variables $\q^{\mu}_\imath $ and $\teb_{\dot{\mu}}^\imath$
are related to each other by complex conjugation: 
$\overline{\q^{\mu}_\imath}=\teb^{\dot{\mu}\imath}$. 
The structure group is ${\rm SL}(2,{\mathbb C})\times {\rm SU}(2)_R \times {\rm U}(1)_R$,
with $M_{ab}=-M_{ba}$, $J_{ij}=J_{ji}$ and $\mathbb J$ be the corresponding 
Lorentz, ${\rm SU}(2)_R$ and ${\rm U}(1)_R$ generators.
The covariant derivatives 
$\cD_{{A}} =(\cD_{{a}}, \cD_{{\a}}^i,\cDB^\ad_i) 
\equiv (\cD_{{a}}, \cD_{ \underline{\a} }, \cDB^{\underline{\ad}})$ 
have the form 
\bea
\cD_{A}&=&E_{A}+
\hf \,\O_{A}{}^{bc}\,M_{bc}+
\Phi_{A}^{~\,kl}\,J_{kl}
+ \ri \,\Phi_{A}\,{\mathbb J} \non\\
&=&E_{A}~+~\O_{A}{}^{\b\g}\,M_{\b\g}
+{\O}_{A}{}^{\bd\gd}\,\bar{M}_{\bd\gd}
+\Phi^{~\,kl}_{A}\,J_{kl}
+\ri \, \Phi_{A}\,{\mathbb J}~.
\label{CovDev}
\eea
Here $E_{{A}}= E_{{A}}{}^{{M}} \pa_{{M}}$ is the supervielbein, 
with $\pa_{{M}}= \pa/ \pa z^{{M}}$,
$\O_{{A}}{}^{bc}$ is the Lorentz connection, 
 $\Phi_{{A}}{}^{kl}$ and $\Phi_{{A}}$ are  
 the ${\rm SU}(2)_R$ and ${\rm U}(1)_R$ connections, respectively.

The Lorentz generators with vector indices ($M_{ab}$) and spinor indices
($M_{\a\b}=M_{\b\a}$ and ${\bar M}_{\ad\bd}={\bar M}_{\bd\ad}$) are related to each other 
by the standard rule:
$$
M_{ab}=(\s_{ab})^{\a\b}M_{\a\b}-(\tilde{\s}_{ab})^{\ad\bd}\bar{M}_{\ad\bd}~,~~~
M_{\a\b}=\hf(\s^{ab})_{\a\b}M_{ab}~,~~~
\bar{M}_{\ad\bd}=-\hf(\tilde{\s}^{ab})_{\ad\bd}M_{ab}~.
$$ 
The generators of the structure group
act on the spinor covariant derivatives as follows:\footnote{The 
(anti)symmetrization of $n$ indices 
is defined to include a factor of $(n!)^{-1}$.}
\bea
{[}M_{\a\b},\cD_{\g}^i{]}
&=&\ve_{\g(\a}\cD^i_{\b)}~,\qquad
{[}\bar{M}_{\ad\bd},\cDB_{\gd}^i{]}=\ve_{\gd(\ad}\cDB^i_{\bd)}~, \non \\
{[}J_{kl},\cD_{\a}^i{]}
&=&-\d^i_{(k}\cD_{\a l)}~,
\qquad
{[}J_{kl},\cDB^{\ad}_i{]}
=-\ve_{i(k}\cDB^\ad_{l)}~, \non \\
{[}{\mathbb J},\cD_{\a}^i{]} &=&\cD_{\a}^i~,\qquad  \qquad ~~\,
{[}{\mathbb J},\cDB^{\ad}_i{]}~
=\,-\cDB^{\ad}_i~,
\label{generators}
\eea
Our notation and conventions correspond to \cite{BK}.

The spinor covariant derivatives obey the algebra
\begin{subequations}\label{eq_algebra}
\bea
\{\cD_\a^i,\cD_\b^j\}&=&
4S^{ij}M_{\a\b}
+2\ve^{ij}\ve_{\a\b}Y^{\g\d}M_{\g\d}
+2\ve^{ij}\ve_{\a\b}\bar{W}^{\gd\dd}{\bar M}_{\gd\dd}
\non\\
&&
+2 \ve_{\a\b}\ve^{ij}S^{kl}J_{kl}
+4 Y_{\a\b}J^{ij}~,
\label{a-c1}\\
%%%%%%%%%%%%%%%%%%%%%%%%%%%%%%%%%%%%%%%%%%%%%
\{\cD_\a^i,\cDB^\bd_j\}&=&
-2\ri\d^i_j(\s^c)_\a{}^\bd\cD_c
+4 \big( \d^{i}_{j}G^{\d\bd}  +\ri G^{\d\bd}{}^{i}{}_{j} \big) M_{\a\d}
+4\big( \d^{i}_{j}G_{\a\gd}
+\ri G_{\a\gd}{}^{i}{}_{j}\big) {\bar M}^{\gd\bd} ~~~~~~
\non\\
&&
+8 G_\a{}^\bd J^{i}{}_{j}
-4\ri\d^i_jG_{\a}{}^\bd{}^{kl}J_{kl}
-2\big( \d^i_jG_{\a}{}^{\bd}
+\ri G_{\a}{}^{\bd}{}^i{}_j\big) {\mathbb J} ~, 
\label{a-c2}
\eea
\end{subequations}
Commutators involving the vector derivative may be worked out
using the Bianchi identity and are summarized in \cite{KLRT-M2}.
The conventions for the Lorentz generators $M_{\alpha \beta}$
and $\bar M_{\dalpha \dbeta}$ as well as the $SU(2)_R$ and $U(1)_R$
generators $J_{ij}$ and $\mathbb J$ are also given in \cite{KLRT-M2}.
Here the dimension-1 components of the torsion
obey the symmetry properties  
\bea
S^{ij}=S^{ji}~, \qquad Y_{\a\b}=Y_{\b\a}~, 
\qquad W_{\a\b}=W_{\b\a}~, \qquad G_{\a\ad}{}^{ij}=G_{\a\ad}{}^{ji}
\eea
and the reality conditions
\bea
\overline{S^{ij}} =  \bar{S}_{ij}~,\quad
\overline{W_{\a\b}} = \bar{W}_{\ad\bd}~,\quad
\overline{Y_{\a\b}} = \bar{Y}_{\ad\bd}~,\quad
\overline{G_{\b\ad}} = G_{\a\bd}~,\quad
\overline{G_{\b\ad}{}^{ij}} = ~G_{\a\bd}{}_{ij}.
\eea
The dimension-3/2 Bianchi identities are:
\begin{subequations}\label{eq_dim3.5bianchi}
\bea
\cD_{\a}^{(i}S^{jk)}&=&0~, \qquad 
\cDB_{\ad}^{(i}S^{jk)} = \ri\cD^{\b (i}G_{\b\ad}{}^{jk)}~,
\label{BI-3/2-1}
 \\
\cD_\a^i\bar{W}_{\bd\gd}&=&0~,\\
\cD_{(\a}^{i}Y_{\b\g)}&=&0~, \qquad 
\cD_{\a}^{i}S_{ij}+\cD^{\b}_{j}Y_{\b\a}=0~, \\
\cD_{(\a}^{(i}G_{\b)\bd}{}^{jk)}&=&0~, \\
\cD_\a^iG_{\b\bd}&=&
- \frac{1}{ 4}\cDB_\bd^iY_{\a\b}
+ \frac{1}{ 12}\ve_{\a\b}\cDB_{\bd j}S^{ij}
- \frac{1}{ 4}\ve_{\a\b}\cDB^{\gd i}\bar{W}_{\gd\bd}
- \frac{\ri }{ 3}\ve_{\a\b}\cD^{\g}_j G_{\g \bd}{}^{ij}~.
\eea
\end{subequations}

This structure is invariant under super-Weyl transformations involving
a real unconstrained parameter $\cU$.
The spinor covariant derivatives transform as
\begin{subequations}
\begin{align}
\cD'_\alpha{}^i &= e^{\cUnew/2} \left(\cD_\alpha{}^i + 2 \cD^{\beta i} \cUnew M_{\beta \alpha}
     - \frac{1}{2} \cD_\alpha{}^i \cUnew \mathbb J + 2\cD_\alpha{}^j \cUnew J_j{}^i\right) \\
\bar\cD'^\dalpha{}_i &= e^{\cUnew/2} \left(\bar \cD^\dalpha{}_i
     - 2 \bar \cD_{\dbeta i} \cUnew \bar M^{\dbeta \dalpha}
     + \frac{1}{2} \bar \cD^\dalpha{}_i \cUnew \mathbb J - 2 \bar \cD^\dalpha{}_j \cUnew J^j{}_i\right)
\end{align}
\end{subequations}
The corresponding torsion superfields are given by
\begin{subequations}
\begin{align}
S'^{ij} &= e^{\cUnew} S^{ij} + \frac{1}{4} e^{3\cUnew} \cD^{ij} e^{-2 \cUnew} \\
Y'_{\alpha \beta} &= e^{\cUnew} Y_{\alpha \beta} -\frac{1}{4} e^{-\cUnew} \cD_{\alpha \beta} e^{2\cUnew} \\
G'_{\alpha \dalpha} &= e^{\cUnew} \cG_{\alpha \dalpha}
     -\frac{1}{16} e^{-\cUnew} [\cD_\alpha{}^k, \bar \cD_{\dalpha k}] e^{2 \cUnew} \\
G'_{\alpha \dalpha}{}^{ij} &= e^{\cUnew} \cG_{\alpha \dalpha}{}^{ij}
     + \frac{\ri}{4} e^{\cUnew} [\cD_\alpha{}^{(i}, \bar \cD_\dalpha{}^{j)}] \cUnew \\
W'_{\alpha \beta} &= e^{\cUnew} W_{\alpha \beta}
\end{align}
\end{subequations}
A conformally primary superfield $\Psi$ of weight $\Delta$ is defined to
transform as
\begin{align}
\Psi' = e^{\cUnew \Delta} \Psi.
\end{align}

Actions in $\cN=2$ supergravity may be constructed from integrals
over the full superspace
\begin{align}
\int \rd^4 x \,{\rm d}^4\q \,{\rm d}^4{\bar \q}
\,E\, \cL
\end{align}
or integrals over a chiral subspace
\begin{align}
\int \rd^4 x \,{\rm d}^4\q \,
\,\cE\, \cL_c ~, \qquad {\bar \cD}^\ad_i \cL_c =0
\end{align}
with $\cE$ the chiral density.
Just as in $\cN=1$ superspace, actions of the former type
may be rewritten as the latter using a covariant chiral projection
operator $\bar \D$ \cite{Muller},
\begin{align}
\int \rd^4 x \,{\rm d}^4\q \,{\rm d}^4{\bar \q}
\,E\, \cL = \int \rd^4 x \,{\rm d}^4\q \,
\,\cE\, \bar\D \cL_c~.
\label{A.12}
\end{align}
The  covariant chiral projection operator is defined as 
\bea
\bar{\D}
&=&\frac{1}{96} \Big((\cDB^{ij}+16\bar{S}^{ij})\cDB_{ij}
-(\cDB^{\ad\bd}-16\bar{Y}^{\ad\bd})\cDB_{\ad\bd} \Big)
\non\\
&=&\frac{1}{96} \Big(\cDB_{ij}(\cDB^{ij}+16\bar{S}^{ij})
-\cDB_{\ad\bd}(\cDB^{\ad\bd}-16\bar{Y}^{\ad\bd}) \Big)~.
\label{chiral-pr}
\eea
Its fundamental property  is that $\bar{\D} U$ is covariantly chiral,
for any scalar, isoscalar and U$(1)_R$-neutral superfield $U(z)$,
\be
{\bar \cD}^{\ad}_i \bar{\D} U =0~.
\ee
A detailed derivation of the relation (\ref{A.12}) can be found in \cite{KT-M}.

\section{$\cN=1$ superconformal geometry}\label{Appendix B}
We give here a summary of the superspace geometry of $\N=1$ conformal
supergravity with the structure group $\rm SL(2,\mathbb C) \times \rm U(1)_R$.
This formulation appeared originally in \cite{Howe} and was elaborated upon
in \cite{GGRS}. Our conventions for generators essentially match those appearing
in Appendix \ref{Appendix A}.

The covariant derivatives have the form
\bea
\nabla_{A} =(\nabla_a , \nabla_\a , {\bar \nabla}^\ad) &=&E_{A}+
\hf \,\O_{A}{}^{bc}\,M_{bc}
+ \ri \,\Phi_{A}\,\hat {\mathbb J}
\eea
and obey the algebra
\begin{subequations}
\begin{align}
\{\nabla_\alpha, \nabla_\beta\} &= -4 \bar R \,M_{\alpha \beta} ~,\\
\{\nabla_\alpha, \bar\nabla_\dalpha\} &= -2\ri \nabla_{\alpha \dalpha} ~,\\
[\nabla_\beta, \nabla_{\alpha \dalpha}] &=
     - \ri \eps_{\beta \alpha} G_{\gamma \dalpha} \,\nabla^\gamma
     + \ri \eps_{\beta \alpha} \bar R \,\nabla_\dalpha
     - \ri \eps_{\beta \alpha} \nabla_\delta G_{\gamma \dalpha} \,M^{\gamma \delta}
     + \ri \bar\nabla_\dalpha \bar R \,M_{\beta \alpha}
     \eol & \quad
     + 2\ri \eps_{\beta \alpha} \bar W_{\dalpha \dbeta \dgamma} \,\bar M^{\dbeta \gamma}
     - \frac{\ri}{3} \eps_{\beta \alpha} \bar X^\dbeta \bar M_{\dbeta \dalpha}
     - \frac{\ri}{2} \eps_{\beta \alpha} \bar X_\dalpha \,\hat {\mathbb J}~.
\end{align}
\end{subequations}
The remaining vector commutator can be calculated using the Bianchi identity.
The $\rm U(1)_R$ generator $\hat{\mathbb J } $ is defined by 
\bea
\left[ \hat {\mathbb J} , \nabla_\a \right] = \nabla_\a~,  \qquad 
\left[ \hat {\mathbb J} ,  {\bar \nabla}_\dalpha \right] =- {\bar \nabla}_\ad~.
\label{B.3}
\eea
The conventional geometry \cite{GGRS, BK} with structure group $\rm SL(2,{\mathbb C})$ 
can be recovered by degauging the $\rm U(1)_R$ and performing a super-Weyl
transformation to set $X_\alpha = 0$.

The components of the torsion are constrained by the Bianchi identities
\begin{subequations}
\begin{gather} 
\bar\nabla_\dalpha R = \bar\nabla_\dalpha W_{\alpha \beta \gamma} = 0 ~,\\
X_\alpha = \nabla_\alpha R - \bar\nabla^\dalpha G_{\alpha \dalpha}~, \qquad
\nabla^\alpha X_\alpha = \bar\nabla_\dalpha \bar X^\dalpha ~,\\
\nabla^\gamma W_{\gamma \beta \alpha} = -\frac{1}{3} \nabla_{(\beta} X_{\alpha)}
     + \ri \nabla_{(\beta}{}^\dgamma G_{\alpha) \dgamma}~.
\end{gather}
\end{subequations}

This structure is invariant under super-Weyl transformations involving a real
unconstrained parameter $U$.
The $\N=1$ covariant derivatives transform as
\begin{subequations}
\begin{align}
\nabla'_\alpha &= e^{\Unew/2} \left(\nabla_\alpha + 2 \nabla^{\beta} \Unew M_{\beta \alpha}
     - \frac{3}{2} \nabla_\alpha \Unew \hat {\mathbb J} \right) ~,\\
\bar\nabla'^\dalpha &= e^{\Unew/2} \left(\bar \nabla^\dalpha
     - 2 \bar \nabla_{\dbeta} \Unew \bar M^{\dbeta \dalpha}
     + \frac{3}{2} \bar \nabla^\dalpha \Unew \hat {\mathbb J} \right)~,
\end{align}
\end{subequations}
and the $\N=1$ torsion superfields transform as
\begin{subequations}
\begin{align}
R' &= e^{\Unew} R -\frac{1}{4} e^{3\Unew} \bar \nabla^2 e^{-2\Unew} ~,\\
G'_{\alpha \dalpha} &= e^{\Unew} G_{\alpha \dalpha}
     + [\nabla_\alpha, \bar \nabla_\dalpha] e^{\Unew} ~,\\
X_\alpha' &= e^{3\Unew/2} X_\alpha - \frac{3}{2} (\bar\nabla^2 - 4 R) \nabla_\alpha \Unew ~,\\
W'_{\alpha \beta \gamma} &= e^{3\Unew/2} W_{\alpha \beta \gamma}~.
\end{align}
\end{subequations}
A conformally primary superfield $\Psi$ of dimension $\Delta$ transforms as
\begin{align}
\Psi' = e^{\Unew \Delta} \Psi~.
\end{align}

Actions in $\N=1$ supergravity may be constructed by integrals over 
the full superspace
\begin{align}
\int \rd^4x\, \rd^2\theta\, \rd^2\bar\theta\, E \,\cL
\end{align}
or integrals over the chiral subspace
\begin{align}
\int \rd^4x\, \rd^2\theta\, \rd^2\bar\theta\, \cE \,\cL_c~, \qquad \bar\nabla_\dalpha \cL_c = 0~.
\end{align}
Actions of the former type may be rewritten as the latter via
\begin{align}
\int \rd^4x\, \rd^2\theta\, \rd^2\bar\theta\, E \,\cL = 
     -\frac{1}{4} \int \rd^4x\, \rd^2\theta\, \cE \,(\bar \nabla^2 - 4 R) \cL~,
\end{align}
and vice versa
\begin{align}
\int \rd^4x\, \rd^2\theta\, \cE \, \cL_c
     = \int \rd^4x\, \rd^2\theta\, \rd^2\bar\theta\, \frac{E}{R} \,\cL_c~.
\end{align}

\section{Improved tensor compensator}
Within the projective superspace formulation given in \cite{KLRT-M1, KLRT-M2},
the $\N=2$ improved tensor compensator
action\footnote{The $\N=1$ improved tensor action was constructed in
\cite{deWR}.} has the form \cite{K-dual08}
\begin{align}
S_{\rm tensor} &= \frac{1}{2\pi}\oint_C  v^i \rd v_i
     \int \rd^4 x \,{\rm d}^4\theta \,{\rm d}^4{\btheta}
     \,E\, \frac{\cW \bar\cW}{(\Sigma^{++})^2}\,
     {\cG}^{++} \ln \frac{{\cG}^{++}}{{\rm i}\breve{ \Upsilon}^{+}  \Upsilon^{+}}~.
\end{align}
Here $\cW$ is the vector compensator and $\Upsilon^+$ is an auxiliary 
arctic multiplet. Although the action appears to depend on the
choice of both $\cW$ and $\Upsilon^+$, one can show \cite{KLRT-M2,K-dual08,KT-M}
that it
is independent of these two fields.

Using the techniques reviewed in \cite{n2_sugra_tensor}, this action can be rewritten
in the form
\begin{align}
S_{\rm tensor} &= \int \rd^4x\, \rd^4\theta\, \cE \Psi \mathbb W + \textrm{c.c.}~,
\end{align}
where
\begin{align}
\mathbb W := \frac{1}{8\pi}\oint_C  v^i \rd v_i \Big((\bar \cD^-)^2 + 4 \bar S^{--}\Big) 
     \ln \frac{{\cG}^{++}}{{\rm i}\breve{ \Upsilon}^{+}  \Upsilon^{+}}
\end{align}
is a reduced chiral superfield by construction. This property is necessary so
that the action is invariant under gauge transformations of the prepotential
$\Psi$ \eqref{eq_PsiGauge}.
The contour integral may be evaluated \cite{n2_sugra_tensor} to give
\begin{align}
\mathbb W = -\frac{\cG}{8} (\cDB_{ij} + 4 \bar S_{ij}) \left(\frac{\cG^{ij}}{\cG^2} \right).
\end{align}
In this form the reduced chirality property is far from obvious and, in fact,
is a tedious calculation to demonstrate. The superfield $\cW$ was first constructed
at the component level in \cite{deWPV}.

In addition the improved tensor action, we have also a coupling between the
tensor multiplet and the vector multiplet which generates the cosmological
constant. In projective superspace, this coupling is written
\begin{align}
S_{\rm cosm} &= - \frac{\xi}{2\pi}\oint_C  v^i \rd v_i
     \int \rd^4 x \,{\rm d}^4\theta \,{\rm d}^4{\btheta}
     \,E\,\frac{\cW \bar \cW}{ (\Sigma^{++} )^2 }\,
     {\cG}^{++} \cV~,
\end{align}
where $\cV$ is the real weight-zero tropical prepotential for $\cW$ \cite{KT-M-ads}
\begin{align}\label{eq_Wprojective}
\cW &= \frac{1}{8\pi} \oint_C  v^i \rd v_i \left((\BCD^-)^2 + 4 \bar S^{--}\right) \cV.
\end{align}
It may be reformulated as a chiral superspace integral
\begin{align}
S_{\rm cosm} &= - \xi \int \rd^4x\, \rd^4\theta\, \cE\, \Psi \cW + \textrm{c.c.}
\end{align}
or as a full superspace integral
\begin{align}
S_{\rm cosm} &= - \xi \int \rd^4x\, \rd^4\theta\, \rd^4\btheta\, E\, \cG^{ij} \cV_{ij}~,
\end{align}
where $\cV^{ij}$ is the unconstrained Mezincescu prepotential \eqref{eq_WMezin}.
The differing forms \eqref{eq_Wprojective} and \eqref{eq_WMezin} are
related and one can derive the latter from the former. The procedure is
described in Appendix E of \cite{n2_sugra_tensor}.

\section{Details for derivation of $\N=2$ action}\label{N2_action_details}
We review in this appendix some details involving the derivation of the linearized
$\N=2$ action \eqref{eq_N2action}. It is useful to isolate it into four sets of terms:
\begin{subequations}
\begin{align}
S^{(2)} &= S_{W}^{(2)} + S_{G}^{(2)} + S_{\rm cosm}^{(2)} + S^{(2)}_{HH} ~,\\
S_{W}^{(2)} &= -\frac{1}{2} \int \rd^4x\, \rd^4\theta\, \cE\, \qW \qW
     - \int \rd^4x\, \rd^4\theta\, \rd^4\bar\theta \, E \,(\bar w \qW \qH + w \bar \qW \qH)~, \\
S_{G}^{(2)} &= \int \rd^4x\, \rd^4\theta\, \cE\, \qPsi \hat{\mathbb W} + \textrm{c.c.}
     + \frac{1}{2g} \int \rd^4x\, \rd^4\theta\, \rd^4\bar\theta \,E \, g_{ij} \qG^{ij} \qH~, \\
S_{\rm cosm}^{(2)} &= -\xi \int \rd^4x\, \rd^4\theta\, \cE \,\qPsi \qW + \textrm{c.c.}
\end{align}
\end{subequations}
$S_W^{(2)}$ and $S_G^{(2)}$ are those terms involving either the vector or
tensor compensators and $S_{\rm cosm}^{(2)}$ consists of the terms involving
both the vector and tensor compensators which arises from the cosmological
term. These are all easy to determine from generalizing the flat space result
and expanding to second order the Noether coupling of $\qH$, as discussed in
\cite{n2_sugra}. The terms $S_{HH}^{(2)}$ are all terms second order in
$\qH$. These can be determined by requiring that the entire action is
gauge invariant. We work with constant background values $w$ and $g^{ij}$
for the compensators for simplicity, but the result will hold for covariantly
constant $\cW$ and $\cG^{ij}$.

The gauge variation of $S_W^{(2)}$ is
\begin{align}
\delta S_{W}^{(2)}
     &= \int \rd^4x\, \rd^4\theta\, \rd^4\bar\theta \,E\, \Bigg\{-4 \bar w \qW \bar S^{ij} \qOmega_{ij}
     + \bar w^2 \qH \bar\Delta (\cD^{ij} + 4 S^{ij}) \qOmega_{ij}
     \eol & \quad
     + \bar w w \qH \bar\Delta (\cD^{ij} + 4 S^{ij}) \bar\qOmega_{ij}
     + \textrm{c.c.}
     \Bigg\}     
\end{align}
and the variation of the cosmological term is
\begin{align}
\delta S_{\rm cosm}^{(2)} = \xi \int \rd^4x\, \rd^4\theta\, \rd^4\bar\theta \,E
     \Bigg\{-4\,\bar\qOmega^{ij} g_{ij} \qW
     + \qPsi (\cD^{ij} + 4 S^{ij})(\bar w \qOmega_{ij} + w \bar\qOmega_{ij})
     + \textrm{c.c.} \Bigg\}~.
\end{align}
Adding these two together and using the AdS relation $S_{ij} = -\xi g_{ij} / w$ gives
\begin{align}
\delta S_{W}^{(2)} + \delta S_{\rm cosm}^{(2)}
     &= \int \rd^4x\, \rd^4\theta\, \rd^4\bar\theta E \Bigg\{
     4\xi \qG^{ij} (\bar w \qOmega_{ij} + w \bar\qOmega_{ij})
     \eol & \quad
     + \bar w^2 \qH \bar\Delta (\cD^{ij} + 4 S^{ij}) \qOmega_{ij}
     + \bar w w \qH \bar\Delta (\cD^{ij} + 4 S^{ij}) \bar\qOmega_{ij} \Bigg\}~.
\end{align}
This depends only on the tensor compensator $\qG^{ij}$
and the supergravity prepotential $\qH$, and so we should be able to cancel it
by terms involving only these prepotentials.

Next we calculate the gauge variation of $S_G^{(2)}$. The analysis
of this for the kinetic term is simplest in projective superspace,
where we observe that
\begin{align}
\int \rd^4x\, \rd^4\theta\, \cE\, \qPsi \hat{\mathbb W} + \textrm{c.c.} =
     \frac{1}{2\pi} \oint v^i \rd v_i \int \rd^4x\, \rd^4\theta\, \rd^4\bar\theta 
     E \frac{\cW \bar \cW}{(\Sigma^{++})^2}
     \frac{ \qG^{++} \qG^{++}}{2 g^{++}}~.
\end{align}
The contour integral variation is given by
\begin{align}
\frac{1}{2\pi} \oint v^i \rd v_i \int \rd^4x\, \rd^4\theta\, \rd^4\bar\theta
     E \frac{\cW \bar \cW}{(\Sigma^{++})^2}
     \delta \qG^{++} \frac{\qG^{++}}{g^{++}}~,
\end{align}
which can be rewritten
\begin{align}
\int \rd^4x\, \rd^4\theta\,\cE\, \delta\qPsi\, \hat{\mathbb W} + \textrm{c.c.}
\end{align}
using eq (5.10) of \cite{n2_sugra_tensor}. This result follows also from
its original chiral form. Using the identity
\begin{align}
\cD_{ij} \qG^{kl} = \frac{1}{3} \delta_{ij}^{kl} \cD_{mn} \qG^{mn}
     + 4 S^{kl} \qG_{ij} - 4 S_{ij} \qG^{kl}
\end{align}
one may show that $\delta S_G^{(2)}$ can be written
\begin{align}
\delta S_G^{(2)}
     &= \int \rd^4x\, \rd^4\theta\, \rd^4\bar\theta \,E\Bigg\{ - 4 \xi \bar w \qOmega_{ij} \qG^{ij}
     + \frac{1}{2g} \qH g^{ij} g_{kl} (\cD_{ij} + 4 S_{ij}) \bar\Delta \bar\qOmega^{kl}
      + \textrm{c.c.} \Bigg\}~.
\end{align}
Adding this to the terms we had before gives
\begin{align}
&\delta S_{G}^{(2)} + \delta S_{W}^{(2)} + \delta S_{\rm cosm}^{(2)} \eol
&=
     \int \rd^4x\, \rd^4\theta\, \rd^4\bar\theta \,E\Bigg\{
     \bar w^2 \qH \bar\Delta (\cD^{ij} + 4 S^{ij}) \qOmega_{ij}
     + \bar w w \qH \bar\Delta (\cD^{ij} + 4 S^{ij}) \bar\qOmega_{ij}
     \eol &\quad
     + \frac{1}{2g} \qH g^{ij} g_{kl} (\cD_{ij} + 4 S_{ij}) \bar\Delta \bar\qOmega^{kl}
     + \textrm{c.c.}\Bigg\}~.
\end{align}
This result, as required, depends only on $\qH$.
$S_{HH}^{(2)}$ must be constructed so that its variation cancels this term.

We construct a solution for $S_{HH}^{(2)}$ by generalizing the Minkowski result
given in \cite{n2_sugra}:
\begin{subequations}
\begin{align}
S_{HH} &= S_{HH.1}+S_{HH.2}+S_{HH.3}+S_{HH.4}~, \\
S_{HH.1} &= -\frac{1}{2} \bar w^2 \int \rd^4x\, \rd^4\theta\, \rd^4\bar\theta \,E \,\qH \bar\Delta \qH
     + \textrm{c.c.}~, \\
S_{HH.2} &= -\frac{1}{64g} \int \rd^4x\, \rd^4\theta\, \rd^4\bar\theta \,E\,
     g_{ij} g_{kl} \qH \cD^{ij} \bar\cD^{kl} \qH ~,\\
S_{HH.3} &= -\frac{g}{32} \int \rd^4x\, \rd^4\theta\, \rd^4\bar\theta \,E\,
     \left(\qH \cD^{ij} \bar\cD_{ij} \qH + \frac{g}{2} \qH \Box \qH \right) ~,\\
S_{HH.4} &= -\frac{g}{8} \int \rd^4x\, \rd^4\theta\, \rd^4\bar\theta \,E\,
     \qH S^{ij} \bar\cD_{ij} \qH + \textrm{c.c.}
\end{align}
\end{subequations}
The term $S_{HH.4}$ is new, having no analogue in a Minkowski background.

It is a straightforward exercise to calculate the variations of the first two terms:
\begin{align}
\delta S_{HH.1} &= \int \rd^4x\, \rd^4\theta\, \rd^4\bar\theta \,E\,
     \Bigg\{- \bar w^2 \qH \bar\Delta (\cD^{ij} + 4 S^{ij}) \qOmega_{ij} 
     - 4 g \bar S^{ij} \qH \Delta \qOmega_{ij} + \textrm{c.c.}\Bigg\} ~,\\
\delta S_{HH.2} &=\int \rd^4x\, \rd^4\theta\, \rd^4\bar\theta \,E\,\Bigg\{
     - \frac{1}{2g} \qH g^{ij} g^{kl} (\bar\cD_{kl} + 4 \bar S_{kl})\Delta \qOmega_{ij}
     \eol & \quad
     + \frac{\ri}{16} g \qH \cD_{\alpha \dalpha} [\cD^{\alpha j}, \bar\cD^\dalpha_j]
     (\cD^{mn} + 4 S^{mn}) \qOmega_{mn}
     + 4 g\qH \bar S^{ij} \Delta \qOmega_{ij}
     \eol & \quad
     - \frac{g}{8} \qH \bar S_{ij} \cD^{ij} (\cD^{mn} + 4 S^{mn}) \qOmega_{mn}
     + \frac{g}{8} \qH S_{ij} \bar\cD^{ij} (\cD^{mn} + 4 S^{mn}) \qOmega_{mn} + \textrm{c.c.}
     \Bigg\}~.
\end{align}
It helps at this point to combine these with the variation of the compensator
terms since numerous cancellations result:
\begin{align}
&\delta S_{G}^{(2)} + \delta S_{W}^{(2)} + \delta S_{FI}^{(2)} + \delta S_{HH.1} + \delta S_{HH.2}
     =
     \eol & \quad
     \int \rd^4x\, \rd^4\theta\, \rd^4\bar\theta \,E\,\Bigg\{
     \frac{\ri}{16} g \qH \cD_{\alpha \dalpha} [\cD^{\alpha j}, \bar\cD^\dalpha_j]
          (\cD^{mn} + 4 S^{mn}) \qOmega_{mn}
     \eol & \quad
     - \frac{g}{8} \qH \bar S_{ij} \cD^{ij} (\cD^{mn} + 4 S^{mn}) \qOmega_{mn}
     + \frac{g}{8} \qH S_{ij} \bar\cD^{ij} (\cD^{mn} + 4 S^{mn}) \qOmega_{mn}
     \eol & \quad
     + g \qH \Delta (\bar\cD^{ij} + \bar S^{ij}) \qOmega_{ij} + \textrm{c.c.} \Bigg\}~.
\end{align}
${}$Finally we can check that
\begin{align}
\delta S_{HH.3}
     &= \int \rd^4x\, \rd^4\theta\, \rd^4\bar\theta \,E\,\Bigg\{
     -2 g \qH \Box (\cD^{mn} + 4 S^{mn})\qOmega_{mn}
     - g \qH \Delta (\bar\cD^{ij} + \bar S^{ij}) \qOmega_{ij}
     \eol & \quad
     - \frac{\ri}{4} g \qH \cD^{\dalpha \alpha} [\cD_\alpha{}^i, \bar\cD_{\dalpha k}]
          (\cD^{kj} + 4 S^{kj}) \qOmega_{ij}
     - \frac{3\ri}{16} g \qH \cD^{\dalpha \alpha} [\cD_\alpha{}^k, \bar\cD_{\dalpha k}]
          (\cD^{ij} + 4 S^{ij}) \qOmega_{ij}
     \eol & \quad
     + \frac{3}{8} g \qH \bar S_{kl} \cD^{kl} (\cD^{ij} + 4 S^{ij}) \qOmega_{ij}
     + \frac{1}{8} g \qH S_{kl} \bar\cD^{kl} (\cD^{ij} + 4 S^{ij}) \qOmega_{ij}
     + \textrm{c.c.} \Bigg\} ~,\\
\delta S_{HH.4} &= \int \rd^4x\, \rd^4\theta\, \rd^4\bar\theta \,E\,\Bigg\{
     - \frac{1}{4} g\qH S^{ij} \bar\cD_{ij} (\cD^{kl} + 4 S^{kl}) \qOmega_{kl}
     \eol & \quad
     - \frac{1}{4} g\qH \bar S^{ij} \cD_{ij} (\cD^{kl} + 4 S^{kl}) \qOmega_{kl} 
     + \textrm{c.c.} \Bigg\}~.
\end{align}
The sum of all these terms is
\begin{align}
\delta S^{(2)} &= \int \rd^4x\, \rd^4\theta\, \rd^4\bar\theta \,E\,\Bigg\{
     -2 g \qH \Box (\cD^{mn} + 4 S^{mn})\qOmega_{mn}
     \eol & \quad
     - \frac{\ri}{4} g \qH \cD^{\dalpha \alpha} [\cD_\alpha{}^i, \bar\cD_{\dalpha k}]
          (\cD^{kj} + 4 S^{kj}) \qOmega_{ij}
     \eol & \quad
     - \frac{\ri}{8} g \qH \cD^{\dalpha \alpha} [\cD_\alpha{}^k, \bar\cD_{\dalpha k}]
          (\cD^{ij} + 4 S^{ij}) \qOmega_{ij}
     + \textrm{c.c.} \Bigg\}
\end{align}
which can be shown to vanish after some complicated algebra.

\section{Details of $\N=1$ reduction}\label{N1_reduction_details}

The $\N=1$ reduction in a Minkowski background was considered in
\cite{n2_sugra}. Most of that work is applicable here since we can
perform a super-Weyl transform to the Minkowski frame, identify the
various $\N=1$ superfields, and then transform back.

We begin by identifying all the $\N=1$ components of the $\N=2$
supergravity multiplet. The AdS frame superfield $\qH$ is related
to the flat frame $\qH_0$ via $\qH = e^{-2\cUnew} \qH_0$, since $\qH$
has super-Weyl weight -2. In the flat frame, the Wess-Zumino gauge
conditions read
\begin{align}\label{eq_WZgaugeHflat}
\qH_0\vert = D_\alpha^{\ul 2} \qH_0\vert = 
     \bar D_\dalpha{}_{\ul 2} \qH_0\vert = (D^{\ul 2})^2 \qH_0\vert = 
     (\bar D_{\ul 2})^2 \qH_0\vert = 0.
\end{align}
It is easy to check that these imply similar-looking conditions in AdS:
\begin{align}
\qH\vert = \cD_\alpha^{\ul 2} \qH\vert = 
     \bar\cD_\dalpha{}_{\ul 2} \qH\vert = (\cD^{\ul 2})^2 \qH\vert = 
     (\bar \cD_{\ul 2})^2 \qH\vert = 0.
\end{align}
Similarly, the flat frame $\N=1$ components
\begin{subequations}
\begin{align}
H_{\alpha \dalpha\, 0} &:= \frac{1}{4} [D_\alpha{}^{\ul 2}, \bar D_{\dalpha}{}_{\ul 2}] \qH_0\vert ~,\\
\Psi_{\alpha\,0} & := \frac{1}{8} (\bar D_{\ul 2})^2 D_\alpha{}^{\ul 2} \qH_0\vert  ~,\\
\hat U_{0} & := \frac{1}{16} D^\alpha{}^{\ul 2} (\bar D_{\ul 2})^2 D_\alpha{}^{\ul 2} \qH_0\vert
     + \frac{1}{12} [D^\alpha, \bar D^\dalpha] H_{\alpha \dalpha\,0}
\end{align}
\end{subequations}
imply in the AdS frame 
\begin{subequations}
\begin{align}
H_{\alpha \dalpha} &:= e^{-\Unew} H_{\alpha \dalpha\, 0} =
     \frac{1}{4} [\cD_\alpha{}^{\ul 2}, \bar \cD_{\dalpha}{}_{\ul 2}] \qH\vert ~,\\
\Psi_{\alpha} &:= e^{-\Unew/2} \Psi_{\alpha\,0} =
     \frac{1}{8} (\bar \cD_{\ul 2})^2 \cD_\alpha{}^{\ul 2} \qH\vert  ~,\\
\hat U &:= \hat U_0 = \frac{1}{16} \cD^\alpha{}^{\ul 2} (\bar \cD_{\ul 2})^2 \cD_\alpha{}^{\ul 2} \qH \vert
     + \frac{1}{12} [\nabla^\alpha, \bar \nabla^\dalpha] H_{\alpha \dalpha}~.
\end{align}
\end{subequations}
We have defined the $\N=1$ components $H_{\alpha \dalpha}$, $\Psi_\alpha$,
and $\hat U$ in AdS by performing an $\N=1$ super-Weyl transformation with
parameter $U$. Since $U = \cU \vert$, it is straightforward to verify the
right hand side of these equations.

The $\N=1$ supergravity gauge transformations in the flat geometry read
\begin{subequations}
\begin{align}
\delta H_{\alpha \dalpha\, 0} &= D_\alpha L_{\dalpha \,0} - \bar D_\dalpha L_{\alpha\,0}~, \\
\delta \Psi_{\alpha\, 0} &= D_\alpha \Omega_0 + \Lambda_{\alpha\,0}~, \\
\delta \hat U_0 &= \rho_0 + \bar\rho_0~.
\end{align}
\end{subequations}
If we choose the gauge parameters $L_\alpha$, $\Omega$, $\Lambda_\alpha$,
and $\rho$ to transform covariantly under super-Weyl transformations,
we recover the AdS frame conditions \eqref{eq_N1gauge}.

For our compensator fields, we had in the flat frame
\begin{align}
\chi_0 := \qW_0 \vert, \qquad W_{\alpha 0}:= \frac{\ri}{2} D_\alpha{}^\2 \qW_0 \vert
\end{align}
for the components of the vector multiplet.
Similarly, the components of the $\N=2$ tensor multiplet $\qG_{ij\,0}$
are given by a chiral scalar $\eta_0$ and a tensor multiplet $L_0$,
\begin{align}
\eta_0 := \qG_{\1\1\, 0}\vert, \quad \bar\eta_0 := \qG_{\2\2\, 0}\vert,
\quad L_0 = -2 \ri \qG_{\1\2\, 0}\vert~.
\end{align}
These generalize quite easily to the corresponding equations in
the AdS frame.

However, in order to derive their transformations under the $\N=1$ supergravity gauge
transformations, we need to first work out their transformations in the flat
geometry. This requires a generalization of the results given in \cite{n2_sugra}
since within the flat geometry the background values $\cG_{ij\, 0}$ and $\cW_0$
are no longer constants. The results are
\begin{subequations}
\begin{align}
\delta \chi_0 &= -\frac{1}{12} \cW_0 \bar D^2 D^\alpha L_{\alpha 0}
     - \frac{1}{4} \bar D^2 (L^\alpha_0 D_\alpha \cW_0)
     - \rho_0 \cW_0 + 2 \ri \Lambda^\alpha_0 \cW_{\alpha\, 0} \\
\delta W_{\alpha 0} &= \frac{1}{4} \bar D^2 D_\alpha \left(
     L^\alpha_0 \cW_{\alpha 0} + L_{\dalpha 0} \bar\cW^\dalpha_0
     + \ri \bar\Omega_0 \cW_0 - \ri \Omega_0 \bar\cW_0 \right)
\end{align}
\end{subequations}
for the components of the $\N=2$ vector multiplet, where
\begin{align}
\cW_{\alpha 0} := \frac{\ri}{2} D_\alpha{}^\2 \cW_0 \vert
\end{align}
is the background value of the $\N=1$ abelian vector field strength.
This vanishes for the choice of $\cW_0$ that we make in this paper,
but we have included it here for full generality.
For the tensor multiplet in the flat frame, we find
\begin{subequations}
\begin{align}
\delta \eta_0 &= -\bar D^2 (\cG_{\ul{12}\, 0} \bar \Omega_0)
     + \cG_{\ul{11}\,0} \rho_0
     - \frac{1}{6} \cG_{\ul{11}\,0} \bar D^2 D^\alpha L_{\alpha 0}
     - \frac{1}{4} \bar D^2 (L^\alpha_0 D_\alpha \cG_{\ul{11}\,0}) \\
\delta L_0 &= -\frac{\ri}{2} D^\alpha \bar D^2 (L_{\alpha 0} \cG_{\ul{12}\,0})
     - \frac{\ri}{2} \bar D_\dalpha D^2 (\bar L^\dalpha_0 \cG_{\ul{12}\,0})
     + \ri D^\alpha (\Lambda_{\alpha 0} \cG_{\ul{11}\,0})
     - \ri \bar D_\dalpha (\bar\Lambda^\dalpha_0 \cG_{\ul{22}\,0}).
\end{align}
\end{subequations}
For the choice of $\cG_{ij\,0}$ we make in this paper, it turns
out that $\cG_{\1\2\, 0}$ vanishes.

Transforming these relations to the AdS frame is straightforward.
It (essentially) involves turning off $\cG_{\1\2\, 0}$ and $\cW_{\alpha \, 0}$,
replacing $\cW_0$ and $\cG_{\1\1\, 0}$ with their constant AdS values
$w$ and $g_{\1\1} = \gamma$, and making the obvious covariantizations of
derivatives everywhere, replacing, for example, $\bar D^2$ with
$\bar\nabla^2 - 4 R$. The results are given in \eqref{eq_N1gaugeCompensators}.

\footnotesize{

}

\end{document}